\documentclass[aps,prd,twocolumn,nofootinbib]{revtex4}
\usepackage{epsfig,graphicx,amssymb,bm}

\newcommand{\simgt}{\lower.5ex\hbox{$\; \buildrel > \over \sim \;$}}
\newcommand{\simlt}{\lower.5ex\hbox{$\; \buildrel < \over \sim \;$}}
\newcommand{\aanda}{A\&A}
\newcommand{\MNRAS}{Mon. Not. Roy. Astr. Soc.}
\newcommand{\ARAA}{Ann. Rev. Astron. Astrophys.}
\newcommand{\apjs}{ApJS}
 \newcommand{\bmf}[1]{\mbox{\boldmath$#1$}}
 \newcommand{\kaco}[1]{\left\langle{#1}\right\rangle}
 \newcommand{\skaco}[1]{\langle{#1}\rangle}

\newcommand{\baredth}{\;\overline{\raise1.0pt\hbox{$'$}\hskip-6pt
\partial}\;}
\newcommand{\edth}{\;\raise1.0pt\hbox{$'$}\hskip-6pt\partial\;}
\begin{document}
\title[Probing dark energy with cluster counts and
cosmic shear power spectra]{
Probing dark energy with cluster counts and
cosmic shear power spectra: including the full covariance}

\author{Masahiro Takada}
\affiliation{Astronomical Institute, Tohoku University,
Sendai, 980-8578, Japan}
\author{Sarah Bridle}
\affiliation{Department of Physics and Astronomy, University College London,
Gower Street, London, WC1E 6BT, UK}

\email{takada@astr.tohoku.ac.jp, sarah.bridle@ucl.ac.uk}

\begin{abstract}
Several dark energy experiments are available from a single large-area
imaging survey, and may be combined to improve
cosmological parameter constraints and/or
test inherent systematics. Two promising
experiments are cosmic shear power spectra
and counts of galaxy clusters. However the two
experiments probe the same cosmic mass density field in large-scale
structure, therefore the combination may be less powerful than first thought.

We investigate the cross-covariance between the
cosmic shear power spectra and the cluster counts based on the halo model
approach, where the cross-covariance arises from the three-point
correlations of the underlying mass density field.
Fully taking into account the cross-covariance as well as
non-Gaussian errors on the lensing power spectrum covariance, we find a
significant cross-correlation between the lensing power spectrum signals
at multipoles $l\sim 10^3$ and the cluster counts containing halos with
masses $M\simgt 10^{14}M_\odot$. Including the cross-covariance for the
combined measurement degrades and in some cases {\em improves} the total
signal-to-noise ratios up to $\sim\pm 20\%$ relative to when the two are
independent.  For cosmological parameter determination, the
cross-covariance has a smaller effect as a result of working in a
multi-dimensional parameter space,
implying that the two observables can
be considered independent to a good approximation.
We also discuss that
cluster count experiments using {\em lensing-selected} mass peaks could be
more complementary to cosmic shear tomography than mass-selected cluster
counts of the corresponding  mass threshold.
Using lensing selected clusters with a realistic usable
detection threshold ($(S/N)_{\rm cluster}\sim 6$ for a ground-based
survey), the uncertainty on each dark energy parameter may be roughly halved
by the combined experiments,
relative to using
the power spectra alone.
\end{abstract}
\maketitle

\section{Introduction}

In recent years great observational progress has been made in measuring the
constituents of the universe (e.g. \cite{spergelea06,coleetal05,tegmarketal06}).
It appears that the universe is currently dominated by an unexpected
component that is causing the universe to accelerate in its expansion.
This component is dubbed ``dark energy''.
Understanding the nature of dark energy is one of most fundamental
questions that remain unresolved with the current cosmological data
sets (e.g. \cite{weinberg89,ratrapeebles88}).
This is now the focus of several planned future surveys
\cite{panstarrs,DES,HSC,LSST,WFMOS,DUNE,SNAP}.

Whether the accelerating expansion is as a consequence of
the cosmological constant,
a new fluid or
a modification to Einstein's gravity,
these future surveys will provide key information.
In addition they will provide a wealth of further cosmological
information, such as constraints on the neutrino mass and the
spectrum of primordial perturbations generated
in the early universe (e.g. \cite{TKF06}).

Combining
several techniques accessible
from different cosmological observables is often a
powerful way to improve constraints on cosmology.
However, care must be taken if the observables are not completely independent.
Two of the most promising methods for constraining the dark energy are
galaxy cluster counts and cosmic shear (e.g. \cite{DETF}).

Clusters of galaxies contain galaxies, hot gas and dark matter in ratio
approximately 1:10:100
 \cite{Sarazin88}.
They are the largest gravitationally bound objects in the
universe and the number of clusters of galaxies has long been
recognized as a powerful probe of cosmology
\cite{lilje92,bahcallc93,hanami93,whiteef93,KitayamaSuto97}.
Counting clusters of galaxies as a function of redshift
allows a combination of structure growth and geometrical information
to be extracted,
thus potentially allowing constraints on the nature of dark energy
\cite{Haimanetal00,Welleretal02,LimaHu04,Wangetal04,MarianBernstein06}.
If cluster masses can be measured accurately then the shape of the
mass function also helps
to break degeneracies \cite{Hu03}.
The distribution of clusters
on the sky (e.g. two-point correlation function) carries additional
information
on dark energy
\cite{MajumdarMohr04,HuHaiman03}.

The bending of light by mass, gravitational lensing, causes images of
distant galaxies to be distorted. These sheared source galaxies are mostly too
weakly distorted for us to measure the effect in single galaxies, but
require surveys containing
a few million galaxies to detect the signal in a statistical way.
This cosmic shear signal has been observed
\cite{vanwaerbekeea00,baconre00,wittmanea00,kaiseretal00}
and used to constrain cosmology (most recently
\cite{jarvisetal05,hoekstraea06,masseyea07,Schimdetal07}).
By using redshift information of source galaxies the evolution of
the dark matter distribution with redshift can be inferred.
Hence, measuring the cosmic shear two-point function as a function of
redshift and separation between pairs of galaxies can be used to constrain
the geometry of the universe as well as the growth of mass clustering.
This method has emerged as one of the most promising to obtain precise
constraints on the nature of dark energy if systematics are well under
control \cite{Hu99,Huterer02,TJ04}.

Future optical imaging surveys suitable for cosmic shear analysis will
also allow the identification of clusters of galaxies.
This could be done either using the colors of the cluster members (e.g.
\cite{koesterea07,gladdersy05})
or using peaks in the gravitational lensing shear field
(e.g. \cite{Miyazakietal02,Wittmanetal06,Miyazakietal07}).
In addition cluster surveys in other wavebands
will overlap with the cosmic shear surveys
allowing detection using X-rays and the thermal Sunyaev--Zel'dovich (SZ)
effect.

Clusters of galaxies produce a large gravitational lensing effect on
distant galaxies, therefore
cluster counts and cosmic shear will not be strictly statistically independent.
The volume surveyed is finite and therefore the number of clusters observed
will not be exactly equal to the average over all universe realizations.
If the number of clusters happens to be higher for a given survey
region,
then the cosmic shear signal is also likely to be higher.
Although the volumes will be large, and thus the deviation is small, this
may amount to a significant uncertainty in the dark energy parameters as
obtained by cluster counts, and dominates the non-Gaussian errors on
the cosmic shear \cite{HuWhite01,coorayhu01,KilbingerSchneider05,TJ07}.

One aspect of this cross-correlation was discussed in \cite{FangHaiman06} and found
to be negligible. However, here we make a full treatment of this
effect using the halo model for non-linear structure formation,
and quantify the resulting change in joint
constraints on the dark energy parameters.

The structure of our paper is as follows. In \S~\ref{prel} we
describe how our observables,
cluster number counts and lensing power spectra, can be expressed
in terms of the background
cosmological
model and the density perturbations.  In
\S~\ref{cov}, we describe a methodology to compute covariances of the
cluster counts and the lensing power spectra, and the cross-covariance
between the two observables.
The detailed
derivations of the covariances are presented in Appendix. In
\S~\ref{results}
we
first study the total signal-to-noise ratios expected for a joint
experiment of the cluster counts and the lensing power spectrum fully
including the cross-covariance predicted from the $\Lambda$CDM
cosmologies. We
then present forecasts for cosmological parameter
determination for the joint experiment, with particular focus on
forecasts for the dark energy parameter constraints.
Finally, we present conclusions and
discussion in \S~\ref{discuss}.

\section{Preliminaries}
\label{prel}

\subsection{A CDM model}

We work in the context of spatially flat cold dark matter models for
structure formation.  The expansion history of the universe is given by
the scale factor $a(t)$ in a homogeneous and isotropic universe (e.g.,
see \cite{Dodelson}).
We describe the Universe
in terms of the matter density $\Omega_{\rm m}$ (the cold dark
matter plus the baryons) and dark energy density $\Omega_{\rm de}$ at
present (in units of the critical density $3H_0^2/(8\pi G)$, where
$H_0=100~ h~{\rm km}~ {\rm s}^{-1}~ {\rm Mpc}^{-1}$ is the Hubble
parameter at present).
In general
the expansion rate, the Hubble parameter, is given
by
\begin{equation}
H^2(a)=H_0^2\left[\Omega_{\rm m}a^{-3}+\Omega_{\rm de}
e^{-3\int^a_1 da' (1+w(a'))/a'}
\right],
\end{equation}
where we have employed the normalization $a(t_0)=1$ today and $w(a)$
specifies the equation of state for dark energy as $w(a)\equiv p_{\rm
de}(a)/\rho_{\rm de}(a)$.  Note that
$\Omega_{\rm m}+\Omega_{\rm de}=1$
and
$w=-1$ corresponds to a
cosmological constant.  The comoving distance $\chi(a)$ from an observer
at $a=1$ to a source at $a$ is expressed in terms of the Hubble
parameter as
\begin{equation}
\chi(a)=\int^1_a\!\!\frac{da'}{H(a')a^{\prime 2}}.
\end{equation}
This gives the distance-redshift relation $\chi(z)$ via the relation
$1+z=1/a$.

Next we need the redshift growth of density perturbations.  In linear
theory
after matter-radiation equality,
all Fourier modes of the mass density perturbation,
$\delta(\bmf{x})(\equiv \delta \rho_m(\bmf{x})/\bar{\rho}_m)$, grow at
the same rate, the growth rate (e.g. see Eq.~10 in \cite{Takada06} for
details).

\subsection{Number counts of galaxy clusters}
\label{cl}

The galaxy cluster observables we will consider in this paper
are the number counts
drawn from a given survey region. Clusters can be
found via their notable observational
properties such as gravitational lensing, member galaxies,
X-ray emission and the SZ effect.
For number counts we simply treat clusters as points; in
other words, we do not care about the distribution of mass within a cluster.
Hence, the number density field of clusters at redshift $z$
can
be expressed as
\begin{equation}
n_{\rm cl}(\bmf{x})=\sum_{i}S(m_i; z)\delta_D^3(\bm{x}-\bm{x}_i),
\label{eqn:ncl}
\end{equation}
where $\delta_D^3(\bmf{x})$ is the three-dimensional Dirac delta
function. The summation runs over halos (the subscript $i$ stands for
the $i$-th halo), and $S(m_i; z)$ denotes the selection function that
discriminates the halos used for the cluster number counts statistic
from other halos.

In this paper, we will consider the following two toy models for the
selection function, to develop intuition for the importance of
cross-correlation between cluster counts and the lensing power spectrum
and to make a comparison between cosmological parameter estimations
derived
from different cluster samples.
Note that throughout this paper we will ignore
uncertainties associated with cluster mass-observable relation,
which could significantly degrade the ability of
cluster counts for constraining cosmological parameters
(e.g. \cite{LimaHu04}). We shall
discuss this issue
 in \S~\ref{sys}.

{\em A mass-limited cluster sample} -- The first toy model we will
consider is a mass-limited cluster sample. For this model, we include
all halos with masses above a given mass threshold:
\begin{eqnarray}
S(m;z)=\left\{
\begin{array}{ll}
1, & \mbox{if $m\ge M_{\rm min}$}\nonumber\\
0, & \mbox{otherwise}.
\end{array}
\right.
\end{eqnarray}
To a zero-th order approximation, the mass-limited selection may
 mimic a cluster sample derived from a flux-limited survey
of clusters via the SZ effect, as this
effect is free of the surface brightness dimming effect
(e.g. see \citep{CarlstromHolderReese02}).

{\em A lensing-based cluster sample} -- A lensing measurement allows one
to make a reconstruction of the two-dimensional mass distribution
projected along the line of sight \cite{KaiserSquires93}. A high peak in
the mass map provides a strong candidate for a massive cluster (see
\cite{Miyazakietal02,Wittmanetal06,Miyazakietal07}
for an implementation of this method to
actual data). To be more explicit, one can define height or significance for
each peak in the reconstructed mass map using the {\em effective}
signal-to-noise ratio (see \cite{HTY} for details):
\begin{eqnarray}
\left(
\frac{S}{N}\right)_{\rm cluster}\equiv
\frac{\kappa_{\rm cluster}(m,z)}{\sigma_{\rm N}}.
\label{eqn:cluster-wl}
\end{eqnarray}
Here $\kappa_{\rm cluster}$ is the convergence amplitude due to a given
cluster at redshift $z$ and with mass $m$,
and $\sigma_{\rm N}$ is 
the rms fluctuations in $\kappa$ due to the
intrinsic ellipticity noise arising from
a finite number of the background galaxies.
Note that we
assume an NFW profile \cite{NFW} with profile parameters modeled in \cite{TJ03a},
and consider the convergence field smoothed with a Gaussian
filter of angular scale $\theta_s=1'$.
To compute the $(S/N)_{\rm
cluster}$ for a cluster at redshift $z$, we
take into account
the remaining fraction of background galaxies behind the cluster for a
given redshift distribution of whole galaxy population (see
\S~\ref{survey}).
This accounts for the variation of
mean redshift and number density of the
background galaxies
with cluster redshift, which changes both the
signal and the intrinsic noise in Eq.~(\ref{eqn:cluster-wl}).

From the reconstructed mass map, a cluster sample may be constructed by
counting mass peaks with heights above a given threshold, $\nu_{\rm
min}$: the selection function is given by
\begin{eqnarray}
S(m; z)=\left\{
\begin{array}{ll}
1, & \mbox{if $(S/N)_{\rm cluster}\ge \nu_{\rm min}$} \nonumber\\
0, & \mbox{otherwise}.
\end{array}
\right.
\label{eqn:sel-wlsn}
\end{eqnarray}
As carefully investigated in \cite{HTY}, the minimum mass of clusters
detectable with a given threshold varies with cluster redshift; clusters
at medium redshift between observer and a typical source redshift are
most easily detectable, while only more massive clusters can be detected
at redshifts smaller and greater than the medium redshift, as
discussed below.

We will employ the halo model to quantify the statistical properties of
cluster observables. In the halo model approach, we assume that all the
matter is in halos. Following the formulation developed in
\cite{ScherrerBertschinger91,
Scoccimarroetal01,TJ03a,TJ03b} (also see Appendix~\ref{app:cluster}
and \cite{CooraySheth} for a thorough review),
the ensemble average of Eq.~(\ref{eqn:ncl}) can be computed as
\begin{eqnarray}
\bar{n}_{\rm cl}
&\equiv& \kaco{n_{\rm cl}(\bmf{x})}=\kaco{
\sum_{i}
 S(m_i; z)\delta_D^3(\bm{x}-\bm{x}_i)}\nonumber\\
&=& \left\langle{\int\!\!dm~\int\!\!d\bm{x}'
\sum_{i}
 S(m; z)\delta_D^3(\bm{x}-\bm{x}')}\right.
\nonumber\\
&&\times \left.{\delta_D(m-m_i)\delta_D^3(\bm{x}'-\bm{x}_i)
}\right\rangle\nonumber\\
&=&\int\!\!dm~ S(m;z)n(m)\int\!\!d\bm{x}'\delta_D^3(\bm{x}-\bm{x}')
\nonumber\\
&=&\int\!\!dm~ S(m;z)n(m),
\label{eqn:3dcl}
\end{eqnarray}
where $n(m)$ is the halo mass function
corresponding to the redshift considered
and we have used the ensemble
average
$\skaco{\sum_i\delta_D(m-m_i)\delta_D^3(\bm{x}_i-\bm{x}')}=n(m)$.
 Thus,
as expected, the ensemble average of the cluster number density field is
given by the integral of the halo mass function, which does not
depend on the cluster distribution and spatial position.
For the halo mass function, we
employ the Sheth-Tormen fitting formula
\cite{ShethTormen}, modified from
the original Press-Schechter function \cite{PressSchechter}.
Note that
we use parameter values $a=0.75$ and $p=0.3$ in the
formula following the discussion in \cite{HuKravtsov03}. We assume that
the mass function can be applied to dark energy cosmologies by replacing
the growth rate appearing in the formula with that for a dark energy
model \cite{LinderJenkins}.

A more useful quantity often considered in the literature is the total number
counts of clusters available from a given survey, which is obtained by
integrating the three-dimensional number density field over a range of
redshifts surveyed. Cluster redshifts are rather
easily available even from a multicolor imaging
survey alone because their central bright galaxies, or red sequence
galaxies, have secure photometric redshift estimates.
Having these facts in mind we will use as our observable the
angular number density averaged over a survey area and divided
into redshift bins:
\begin{eqnarray}
{\cal N}_{(b)\rm cl}&=&\int\!\!d^2\bm{\theta}~W(\bm{\theta})
\int_0^{\chi_H}\!\!d\chi~ \frac{d^2V}{d\chi d\Omega}
\nonumber\\
&&\hspace{-2em}\times
\sum_i S_{(b)}(m_i; z)
\delta_D(\chi-\chi_i)\delta_D^2(\chi\bm{\theta}-\chi_i\bm{\theta}_i),
\label{eqn:angcl}
\end{eqnarray}
where $W(\bm{\theta})$ is the window function of the survey defined so
that it is normalized as $\int\!\!d^2\bm{\theta}W(\bm{\theta})=1$,
$\chi_H$ is the
distance to the Hubble horizon,
and the comoving volume per
unit comoving distance and unit steradian is given by $d^2V/d\chi
d\Omega=\chi^2$
for a flat universe.  The subscript in the round bracket,
$(b)$, stands for the $b$-th redshift bin for the cluster number counts.
In the
following, we will simply consider the sharp redshift selection function
\begin{eqnarray}
S_{(b)}(m_i; z)\rightarrow \left\{
\begin{array}{ll}
S(m_i), & \mbox{if $z_{(b),{\rm lower}}\le z\le z_{(b), {\rm upper}}$}\\
0, & \mbox{otherwise.}
\end{array}
\right.
\end{eqnarray}
Note that the
redshift $z$ appearing in the argument of $S_{(b)}(m_i; z)$ is related
to the comoving distance $\chi$ via the relation $d\chi = dz/H(z)$.

Using the halo model, the expectation value of the angular number
density can be computed from the ensemble average of
Eq.~(\ref{eqn:angcl}) as
\begin{eqnarray}
N_{(b)}&\equiv& \skaco{{\cal N}_{(b)}}\nonumber\\
&=& \int\!\!d^2\bm{\theta}W(\bm{\theta})
\!\int^{\chi_H}_0\!\!d\chi \frac{d^2V}{d\chi d\Omega}
 \int\!\!dm~ S_{(b)}(m;z)n(m) \nonumber\\
&=& \int^{\chi_H}_0\!\!d\chi~
\frac{d^2V}{d\chi d\Omega}
 \int\!\!dm~ S_{(b)}(m;z)n(m).
 \label{eqn:cc-Nb}
\end{eqnarray}
Thus, the expectation value again does not depend on the cluster
distribution.  The sensitivity of the number density to dark energy
arises from the comoving volume and the mass function $n(m)$
\cite{Haimanetal00}.

\begin{figure}[th]
\centerline{\psfig{file=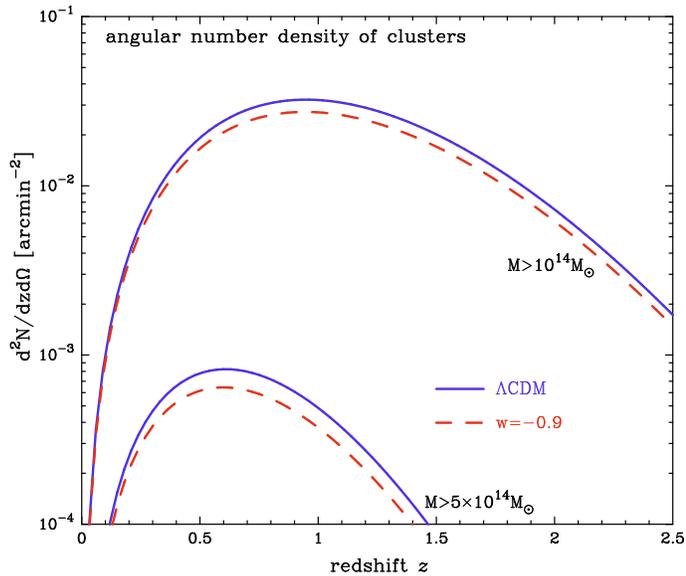, width=9.cm}}
\caption{The average angular number density of halos with masses above a given
 threshold, per unit square arcminute and per unit redshift
 interval. The upper pair
 and lower pair
 of curves are for halos with $M/M_\odot\ge
 10^{14}$ and $5\times 10^{14}$, respectively.
Increasing
 the dark energy
 equation state from $w=-1$ to  $w=-0.9$ decreases the number density, as
 shown by the dashed curves.  } \label{fig:dndz}
\end{figure}
\begin{figure}[th]
\centerline{\psfig{file=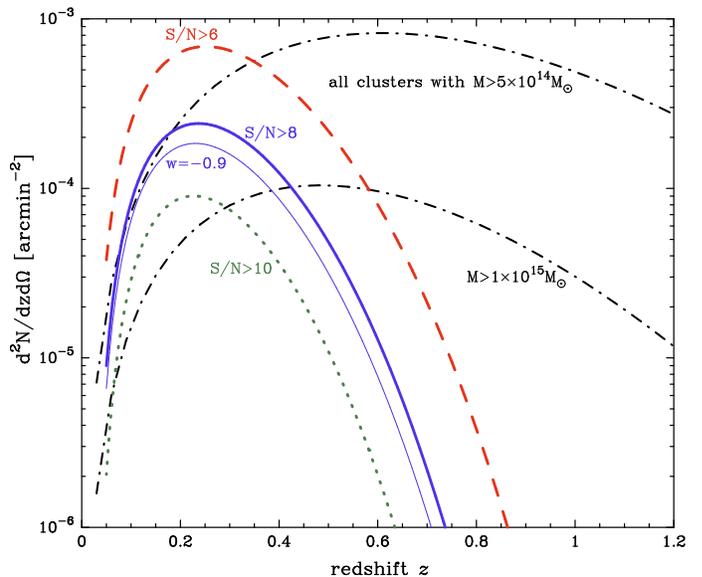, width=9.cm}}
\caption{As in the previous plot, but shown here is the number density
for the lensing-based cluster sample, where clusters having a lensing
signal greater than a given detection threshold are selected in the sample as
described around Eq.~(\ref{eqn:sel-wlsn}). The dashed, solid and dotted
curves show the results for the detection thresholds $(S/N)_{\rm
cluster}\ge 6,8$ and $10$, respectively.  For comparison, the two
dot-dashed curves show the number density for the mass-selected cluster
sample with masses $M/M_\odot \ge5, 10\times 10^{14}$.  Increasing $w_0$
from $w_0=-1.0$ to $w_0=-0.9$ leads to a decrease in the number density
as shown by the thin-solid curve, compared to the bold-solid curve.  The
lensing selected
number densities
peak
at a redshift $z\sim 0.25$, reflecting redshift dependence of the
lensing efficiency function for source galaxies at $z_{\rm s}\sim 1$.
} \label{fig:dndz-wl}
\end{figure}
Fig.~\ref{fig:dndz} shows the average angular
number density of halos with
masses greater than a given threshold, per unit square arcminute and
per unit redshift interval
assuming the fiducial model defined in~\ref{survey}.
Increasing
the dark energy equation of state
from our fiducial model $w=-1.0$ to $w=-0.9$ decreases the number density,
because the change decreases both the comoving volume
$d^2V/d\chi d\Omega$
and the
number density of cluster-scale halos, for a given CMB normalization
of density perturbations.
Comparing the results for mass thresholds
$M_{\rm min}/M_\odot=10^{14}$ and $5\times 10^{14}$ clarifies that a
factor 5 increase in the mass threshold leads to a significant decrease
in the number density, reflecting the mass sensitivity of the halo mass
function in its exponential tail.

In Fig.~\ref{fig:dndz-wl} we present the number density for the
lensing-based cluster sample in which clusters having a lensing signal
greater than a given detection threshold are included in the sample as discussed
around Eq.~(\ref{eqn:sel-wlsn}). Note that to compute the results shown
in this plot we assumed the redshift distribution of galaxies
described in \S~\ref{survey} and the NFW mass profile to model
the cluster lensing. 
In practice
high detection thresholds such as
$(S/N)_{\rm cluster}\simgt 6$
are necessary
in order to make robust
estimates for cluster counts, because contamination of false
peaks due to intrinsic ellipticities or the projection effect
are expected to be low for
such high thresholds
(see \cite{WhiteVanWaerbeke02,HTY} for the details).
Comparing with the number
density for a mass-selected sample shown by the dot-dashed curves,
one can roughly find which mass and redshift
ranges of clusters are probed by the lensing-based cluster sample.
For example, the cluster sample with
lensing signal $(S/N)_{\rm cluster}\ge 10$ contains massive clusters
with masses $M\simgt 10^{15}M_\odot$ over redshift ranges $z\simlt 0.4$,
while only even more massive clusters are included in the sample at the
higher redshifts.  This cluster sample has a narrower redshift coverage
than the simple mass threshold;
all the curves peak at a redshift $z\sim 0.25$. The peak redshift is
mainly attributed to redshift dependence of the lensing efficiency for
source galaxies of $z_s\sim 1$
in our redshift distribution.
A change of $w_0$
from $w_0=-1.0$ to $w_0=-0.9$ leads to a decrease in the number density,
as seen in Fig.~\ref{fig:dndz}.
As before the effect comes partially arises  from the decrease in
comoving volume and the change in the halo mass function.
Unlike the simple mass threshold case, there is now an additional
contribution to the decrease in number density caused by the lower
lensing efficiency and thus lower $S/N$ for a cluster of a given mass
and redshift.

\subsection{Lensing power spectrum with tomography}
\label{lens}

Gravitational shear can be simply related to the lensing convergence:
the weighted mass distribution integrated along the line of sight.
Photometric redshift information on source galaxies allows us to
subdivide galaxies into redshift bins
(we will
discuss possible effects of photometric redshift errors on our
results in \S~\ref{sys}).
This allows more cosmological
information to be extracted,
which is referred to as lensing tomography
(e.g., see
\cite{Mellier99,BartelmannSchneider,Schneider06} for a thorough review, and see
\cite{Hu99,Huterer02,TJ04} for the details of lensing tomography).

In the context of cosmological
gravitational lensing the convergence field with tomographic
information is expressed as a weighted projection of the
three-dimensional mass density fluctuation field:
\begin{equation}
\kappa_{(i)}(\bmf{\theta})=\int_0^{\chi_H}\!\!d\chi W_{(i)g}(\chi)
\delta[\chi, \chi\bmf{\theta}],
\label{eqn:kappa}
\end{equation}
where $\bmf{\theta}$ is the angular position on the sky,
and $W_{(i)g}$ is the
gravitational
lensing weight
function for source galaxies sitting in the $i$-th redshift bin (see
Eq.~(10) in \cite{TJ04} for the definition). Note that, hereafter,
quantities with subscripts in the round bracket such as $(i)$ stands for
those for the $i$-th redshift bin.
To avoid confusion, throughout
this paper we use $i,j$ or $i',j'$ for the lensing power spectrum
redshift bins, and $b,b'$ for the cluster count redshift bins.

The lensing tomographic information
allows us to extract redshift evolution of the lensing weight function
as well as the growth rate of mass clustering.
These
are both sensitive
to dark energy. For example, increasing the equation of state parameter
$w$ from $w=-1$ lowers $W_{(i)g}$ as well as suppressing the growth rate
at lower redshifts.
Therefore when the CMB normalization of density perturbations is
employed, an increase in $w$ decreases the lensing power spectrum
due to both the lower $W_{(i)g}$ and the lower matter power spectrum amplitude.
The sensitivity of lensing observables to the dark energy equation of
state roughly arises equally from the two effects (e.g., see \cite{simpsonb05}).

The cosmic shear fields are measurable only in a statistical way.  The
most conventional methods used in the literature are the shear two-point
correlation function. The Fourier transformed counterpart is the shear
power spectrum.
The convergence
power spectrum is identical to the shear power spectrum but is
easier to work with as it is a scalar.
Using the flat-sky approximation \cite{Limber}, the angular power
spectrum between the convergence fields of redshift bins $i$ and $j$ is
found to be
\begin{equation}
P_{(ij)\kappa}(l)=\int_0^{\chi_H}\!\!d\chi W_{(i)g}(\chi)
W_{(j)g}
(\chi)
\chi^{-2}
P_\delta\!\left(k=\frac{l}{\chi}; \chi\right),
\label{eqn:p_kappa}
\end{equation}
where $P_\delta(k)$ is the three-dimensional mass power spectrum.
We can safely employ the flat-sky approximation for our purpose, because
a most accurate measurement for the lensing power spectrum is available
around multipoles $l\sim 1000$ for a ground-based survey of our interest
(e.g. see Fig.~1 in \cite{Jainetal07}), and the flat-sky approximation
serves as a very good approximation on these small scales \cite{Hu00}.

 For
$l\simgt 100$ the major contribution to $P_{(ij)\kappa}(l)$ comes from
non-linear clustering (e.g., see Fig.~2 in \cite{TJ04}).  We employ the
fitting formula for the non-linear $P_\delta(k)$ proposed in Smith et al.
\cite{Smith}, assuming that it can be applied to dark energy cosmologies
by replacing the growth rate used in the formula with that for a given
dark energy model.  We note in passing that the issue of accurate power
spectra for general dark energy cosmologies still needs to be addressed
carefully (see \cite{HutererTakada,Ma07} for related discussion).
\begin{figure}[th]
\centerline{\psfig{file=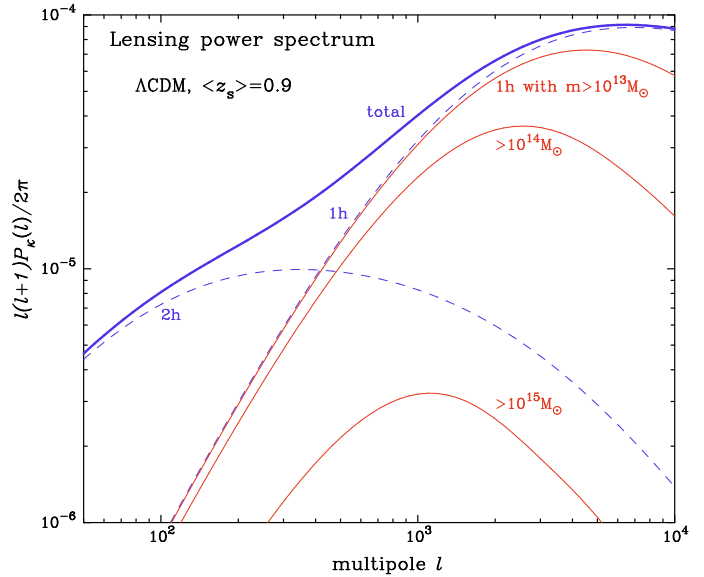, width=9.cm}}
 \caption{A lensing
  power spectrum for the
  non-tomographic case (i.e. one redshift bin) for a $\Lambda$CDM model,
  expected for a ground-based survey that probes galaxies with mean
  redshift $\skaco{z_s}=0.9$. The two thin dashed curves show the 1- and
  2-halo term contributions to the power spectrum, while the bold curve shows the
  total power. The three thin solid curves show the 1-halo term
  contributions obtained when the lensing effects on background galaxies
  due to halos with masses $M/M_\odot\ge 10^{15}, 10^{14}, 10^{13}$ are
included,
respectively. } \label{fig:cl}
\end{figure}
Fig.~\ref{fig:cl} demonstrates how lensing of background galaxies by clusters
contributes to the lensing power spectrum.
Note here that we have employed the halo model developed in
Takada \& Jain \cite{TJ03a,TJ03b}
to compute the mass
power spectrum, although we will use the Smith et al. fitting formula
to compute the lensing power spectrum
in
most parts of this paper
instead.
Briefly, to compute the spectra based on the halo model
approach, we need to model three ingredients:
(i)
the halo mass function
(see also the description below Eq.~[\ref{eqn:3dcl}]);
(ii)
the
profile
for
the
mass distribution around a halo; and
(iii)
the halo bias parameter.

 It is clear that the convergence on
scales $l\simgt 100$ is significantly boosted by the existence of
non-linear structures, halos. In this paper we are especially interested
in using the lensing information inherent in angular scales $l\simlt
3000$ \footnote{At the smaller angular scales $l\simgt 3000$, more
complex uncertainties in non-linear clustering such as the baryonic
effects arise, which need to be addressed more carefully.}
 to constrain dark energy, and
a fair fraction of the power
at scales $l\sim 10^3$, up to $\sim 60\%$ of the total power,
arises from massive halos with $M\simgt
10^{14}M_\odot$. The 1-halo term contribution is given by redshift-space
integral of the halo mass function and halo profiles weighted with
the lensing efficiency.  The results imply that, if massive clusters
with $M\simgt 10^{14}M_\odot$ happen to be less or more populated in a
survey region, amplitudes of the observed lensing power spectrum from
the survey are very likely to be
smaller or greater than expected, respectively.
Therefore, a cross-correlation between the lensing power spectrum and
the cluster counts are intuitively expected, if both of the observables
are measured from the same survey region.

In reality, the observed power spectrum is contaminated by the intrinsic
ellipticity noise.  Assuming that the intrinsic ellipticity distribution
is uncorrelated between different galaxies, the observed power spectrum
between redshift bins $i$ and $j$ can be expressed as
\begin{equation}
P^{\rm obs}_{(ij)\kappa}(l)
=P_{(ij)\kappa}(l)+\delta^K_{ij}\frac{\sigma_\epsilon^2}{\bar{n}_{(i)}}
\label{eqn:psobs}
\end{equation}
where $\sigma_\epsilon$ is the rms of intrinsic ellipticities per
component, and $\bar{n}_{(i)}$ denotes the average number density of
galaxies in the $i$-th redshift bin.  The Kronecker delta function,
$\delta^K_{ij}$, accounts for the fact that the cross-spectra of
different redshift bins ($i\ne j$) are not affected by the shot noise
contamination. We will omit the superscript `obs' when referring to
$P_\kappa^{\rm obs}(l)$ in the following for notational simplicity.

\section{Covariances of lensing power spectrum and cluster observables}
\label{cov}

To estimate a realistic forecast for
cosmological parameter
constraints for a given
survey we have to quantify sources of statistical error on
observables of interest, the cluster number counts and the lensing power
spectrum, and then
 propagate the errors into the parameter forecasts. In
this section, we will
present
the covariance matrices of the observables.

\subsection{Covariances of the cluster number counts}

The cluster observables can be naturally incorporated in the halo model
approach, allowing us to compute the statistical properties
in a straightforward way. In this paper we focus on the
average angular number density of clusters drawn from a survey, also subdivided
into redshift bins as described in \S~\ref{cl}. The covariance between
the average number densities in redshift bins $b$ and $b'$, given by
Eq.~(\ref{eqn:angcl}), is defined as
\begin{eqnarray}
[\bm{C}]^c_{bb'}&\equiv&\skaco{{\cal N}_{(b)}{\cal
 N}_{(b')}}-N_{(b)}N_{(b')}.
\label{eqn:clcov-def}
\end{eqnarray}
Based on the halo model the covariances of the angular number density
 can be derived in Appendix~\ref{appendix:clcov}
(also see
 \cite{HuKravtsov03} for the original derivation)
as
\begin{eqnarray}
[\bm{C}^c]_{bb'}&=&\delta^K_{bb'}\frac{N_{(b)}}{\Omega_{\rm s}}\nonumber\\
&&\hspace{-4em}+\delta^K_{bb'}\int_0^{\chi_H}\!\!\!d\chi \!\left(
\frac{d^2V}{d\chi d\Omega}
     \right)^2\!\!\chi^{-2}\!\!
\left[\int\!\!dm~ n(m)S_{(b)}(m;\chi)b(m)\right]^2\nonumber\\
&&\hspace{-3em}\times
\int\!\!\frac{ldl}{2\pi}P^L_\delta\!\!\left(k=\frac{l}{\chi}; \chi\right)
|\tilde{W}(l\Theta_s)|^2,
\label{eqn:clcov}
\end{eqnarray}
where $b(m)$ is the halo bias parameter (\cite{MoWhite96}; we use the
model derived in \cite{ShethTormen}),
$P^L_{\delta}(k)$
is the linear mass power
spectrum, and $\tilde{W}(x)$ is the Fourier transform of the survey
window function; for this we simply employ
$\tilde{W}(l\Theta_s)=2J_1(l\Theta_s)/(l\Theta_s)$ ($J_1(x)$ is the 1-st
order Bessel function) assuming a circular geometry of the survey
region, $\Omega_{\rm s}=\pi\Theta_s^2$.
In the
following, the tilde symbol is used to denote the Fourier components
of quantities.
 To
derive the covariance (\ref{eqn:clcov}), we have ignored correlations
between the number densities between different redshift bins, which
would be a good approximation for a redshift bin thicker than the
correlation length of the cluster distribution.

The first and second terms in Eq.~(\ref{eqn:clcov})
arise from the 1- and 2-halo terms in the halo model calculation; the
former gives the shot noise due to the imperfect sampling of
fluctuations by a finite number of clusters, while the latter
represents the sampling variance arising from fluctuations of the
cluster distribution due to a finite survey volume. It should be noted
that our formulation allows us to derive the shot noise term without
{\em ad hoc} introducing as often done in the literature
(e.g., see \cite{Haimanetal00}).  The two terms in Eq.~(\ref{eqn:clcov})
depend on sky coverage in slightly different ways\footnote{If a new
integration variable $x=l\Theta_{ s}$ is introduced for the second term
of Eq.~(\ref{eqn:clcov}), one can find the sky coverage dependence is
expressed as
$\propto (1/f_{\rm sky})\int_0^{\infty}\!\!dx~
xP(k=x/\Theta_{\rm s}\chi) |\tilde{W}(x)|^2$, which looks similar to the
$f_{\rm sky}$ dependence of the first term given as
$\propto 1/f_{\rm sky}$.
However,
the $f_{\rm sky}$ dependence could be different via the dependence in
$P(k=x/\Theta_{\rm s}\chi)$ for the 2nd term.}, and the relative
importance depends on the survey area; for a larger survey, the sampling
variance could be more important than the shot noise
\cite{HuKravtsov03}.

\subsection{Covariances of lensing power spectra}

In reality the lensing power spectrum has to be estimated from the
Fourier
or spherical harmonic
coefficients of the observed lensing fields constructed for a
finite survey.
In this paper we assume the flat-sky approximation and thus use Fourier
wavenumbers $\ell$, which are equivalent to spherical harmonic
multipoles $\ell$ in the limit $\ell \gg 1$ \cite{Hu00}.
Because the survey is finite,
an infinite number of Fourier modes
are not available, and rather the discrete Fourier decomposition has to
be constructed in terms of the fundamental mode that is limited by the
survey size; $l_{\rm f}=2\pi/\sqrt{\Omega_{\rm s}}$, where $\Omega_{\rm
s}$ is the survey area. We assume a homogeneous survey
geometry for simplicity and do not consider any complex boundary and/or
masking effects.
The lensing power spectrum of a multipole $l$
is observationally estimated
by averaging over wavenumber direction in an annulus of width
$\Delta l$
\begin{eqnarray}
P_{(ij)\kappa}^{\rm est}(l)&=&\int_{|\bm{l}'|\in l}
\!\!\frac{d^2\bm{l}'}{A(l)}
\tilde{\kappa}_{(i)\bm{l}'}\tilde{\kappa}_{(j)-\bm{l}'},
\label{eqn:psest_maintext}
\end{eqnarray}
where
the integration range is confined to the
Fourier modes of $\bm{l}'$ satisfying the bin condition $l-\Delta l/2\le
l'\le l+\Delta l/2$ and $A(l)$ denotes the integration area in the
Fourier space approximately given by $A(l)\equiv
\int_{|\bm{l}'|\in l}d^2\bm{l}'\approx 2\pi l\Delta l$.
This is discussed in more detail in Appendix~\ref{appendix:wlcov}.

Once an estimator of the lensing power spectrum is defined, it is
straightforward to compute the covariance
\cite{Scoccimarroetal99,coorayhu01} (also see \cite{TJ07} for the
detailed derivation). From Eq.~(\ref{eqn:ps-cov}),
the covariance to
describe the correlation between the lensing power spectra of different
multipoles and redshift bins is given by
\begin{eqnarray}
[\bm{C}^{g}]_{mn} &\equiv& \skaco{P^{\rm est}_{(ij)\kappa}(l)
P^{\rm est}_{(i'j')\kappa}(l')}-P_{(ij)\kappa}(l)P_{(i'j')\kappa}(l')
\nonumber\\
&&\hspace{-5em}=\frac{2\delta^K_{ll'}}{(2l+1)\Delta lf_{\rm sky}}
\left[P_{(ii')\kappa}\!(l)P_{(jj')\kappa}\!(l)+
P_{(ij')\kappa}\!(l)P_{(ji')\kappa}\!(l)
\right]\nonumber\\
&&\hspace{-5em}+\frac{1}{4\pi f_{\rm sky}}
\int_{|\bm{q}|\in l}\!\!\frac{d^2\bm{q}}{A(l)}
\int_{|\bm{q}'|\in l'}\!\!\frac{d^2\bm{q}'}{A(l')}
T_{(iji'j')\kappa}(\bm{q},-\bm{q},\bm{q}',-\bm{q}'),\nonumber\\
\label{eqn:pscov}
\end{eqnarray}
where $f_{\rm sky}$ is the sky coverage ($f_{\rm sky}=\Omega_{\rm
s}/4\pi$) and the lensing trispectrum $T_\kappa$ is defined in terms of
the 3D mass trispectrum $T_\delta$ as
\begin{eqnarray}
T_{(iji'j')\kappa}(\bm{l}_1,\bm{l}_2,\bm{l}_3,\bm{l}_4)
&=&\int_0^{\chi_H}\!\!d\chi W_{(i)g}W_{(j)g}W_{(i')g}W_{(j')g}
\nonumber\\
&&\times \chi^{-6} T_{\delta}(\bm{k}_1,\bm{k}_2,\bm{k}_3,\bm{k}_4
; \chi),
\label{eqn:lenstrisp}
\end{eqnarray}
with $\bm{k}_i=\bm{l}_i/\chi$. Note that the power spectra
$P_{(ij)\kappa}$ appearing on the r.h.s. of Eq.~(\ref{eqn:pscov}) are the
observed spectra given in
Eq.~(\ref{eqn:psobs}), and therefore include the intrinsic ellipticity
noise.  The indices $m,n$ denote elements in the lensing
power spectrum covariance and run over the multipole bins and redshift
bins. For tomography with $n_z$ redshift bins, there are
$n_z(n_z+1)/2$ different spectra available at each multipole. Hence, if
assuming $n_l$ multipole bins, the indices $m,n$ run as
$m,n=1,2,\dots,n_ln_z(n_z+1)/2$. In most parts of this paper we adopt $100$
multipole bins logarithmically spaced,
which are sufficient
to capture all the relevant features in the lensing power spectrum.
For example, for tomography with 3 redshift bins,
the covariance matrix $\bm{C}^{g}$ has dimension of
$600\times600$  for $n_l=100$.

The first term of the covariance matrix (second line of
Eq.~[\ref{eqn:pscov}])
represents the
Gaussian error contribution ensuring that the two power spectra of
different multipoles are uncorrelated via $\delta^K_{ll'}$, while the
second term gives the non-Gaussian errors to describe correlation
between the different power spectra.  The two terms both scale with sky
coverage as $\propto 1/f_{\rm sky}$. Note that the non-Gaussian term
does not depend on the multipole bin width $\Delta l$ because of
$\int\!\!d^2\bm{q}/A(l)\approx 1$, and taking a wider bin only reduces the
Gaussian contribution or equivalently enhances the relative importance
of the non-Gaussian contribution.
Naturally, however, the signal-to-noise ratio and parameter forecasts
we will show below do not depend on the multipole bin width
if the bin width is not very coarse
(see
\cite{TJ07} for the details).

We employ a further simplification to make quick computations of the
lensing covariance matrices. We use the halo model approach to compute
the lensing covariance matrices. We know that most of the signal in the
power spectrum comes from small angular scales at $l\sim 10^3$ to which
the 1-halo term provides dominant contribution as shown in
Fig.~\ref{fig:cl}. In addition, the non-Gaussian errors are important
only at small angular scales.  For these reasons, we only include the
1-halo term contribution to the lensing trispectrum to compute the
non-Gaussian errors. Although the trispectrum generally
depends on four vectors in
the Fourier space such as $\bm{l}_1,\bm{l}_2,\bm{l}_3$ and $\bm{l}_4$,
the 1-halo term does not depend on any angle between the vectors, but
rather depends only on the length of each vector;
$T^{1h}(\bm{l}_1,\bm{l}_2,\bm{l}_3,\bm{l}_4)= T^{1h}(l_1,l_2,l_3,l_4)$
(see \cite{TJ03a,TJ03b,TJ07}), reflecting spherical mass distribution
around a halo in a statistical average sense, which does not have any
preferred direction in the Fourier space.  Therefore, the non-Gaussian
term in Eq.~(\ref{eqn:pscov}) can be further simplified as
\begin{eqnarray}
&&\frac{1}{4\pi f_{\rm sky}}
\int_{|\bm{q}|\in l}\!\!\frac{d^2\bm{q}}{A(l)}
\int_{|\bm{q}'|\in l'}\!\!\frac{d^2\bm{q}'}{A(l')}
T_{(iji'j')\kappa}(\bm{q},-\bm{q},\bm{q}',-\bm{q}')
\nonumber \\
&&\hspace{3em}\approx \frac{1}{4\pi f_{\rm
 sky}}T^{1h}_{(iji'j')\kappa}(l,l,l',l').
\end{eqnarray}
where we have assumed that the lensing trispectrum does not
change significantly within the multipole bin, which is a good approximation for the
lensing fields.

\subsection{Cross-covariances of the cluster number counts and
lensing power spectra}

The cluster observables and the weak lensing power spectra probe the
same density fluctuation fields in large-scale structure, if the two
observables are drawn from the same survey region.  As implied in
Fig.~\ref{fig:cl}, a somewhat significant correlation between the two
observables is expected if the small-scale lensing power spectrum is
considered.
We again use the halo model to compute the
cross-covariance. The detailed derivation is described in
Appendix~\ref{appendix:ccwlcov},
and the cross-covariances can be expressed as
\begin{eqnarray}
[\bm{C}^{gc}]_{mb}&\equiv& \skaco{P^{\rm est}_{(ij)\kappa}(l){\cal
 N}_{(b)}}-P_{(ij)\kappa}(l)N_{(b)}\nonumber\\
&&\hspace{-5em}=\frac{1}{\Omega_{\rm s}}
\int_0^{\chi_H}\!\!d\chi \frac{d^2V}{d\chi
 d\Omega}W_{(i)g}W_{(j)g}\chi^{-4}
 B_{(b)c\delta\delta}(k=l/\chi). \nonumber \\
\label{eqn:cpscov}
\end{eqnarray}
Here $B_{(b)c\delta\delta}$ is the 3D bispectrum corresponding to the
three-point function of the
cluster distribution and the two mass density fluctuation fields. The
cross-covariance arises from two contributions of the 3D bispectrum, the
1- and 2-halo terms:
\begin{eqnarray}
B^{1h}_{(b)c\delta\delta}(k; \chi)&=&\int\!\!dm~ n(m)S_{(b)}(m)
\left(\frac{m}{\bar{\rho}_m}\right)^2 \tilde{u}_m^2(k)\nonumber\\
B^{2h}_{(b)c\delta\delta}(k; \chi)&=&2
\left[
\int\!\!dm_1~ n(m_1)b(m_1)\frac{m_1}{\bar{\rho}_m}\tilde{u}_{m_1}(k)
\right]
\nonumber\\
&&\hspace{-6em}\times
\left[
\int\!\!dm_2~ n(m_2)S_{(b)}(m_2)b(m_2)\frac{m_2}{\bar{\rho}_m}\tilde{u}_{m_2}(k)
\right]P^L(k),\nonumber\\
\label{eqn:cov1h-2h}
\end{eqnarray}
where $\tilde{u}_m$ is the Fourier transform of a halo profile for which
we assume an NFW profile \cite{NFW} as explicitly defined in
Eq.~(\ref{eqn:um}).
The cross-covariance arising from
the 1-halo term represents correlation between one cluster, treated as a
point, and
the lensing effects on two different  background galaxies due to the same
cluster. The 2-halo term contribution shows the correlation between one
cluster, the lensing field on a background galaxies around the cluster,
and the lensing field due to another cluster.
Note that the cross-covariance (\ref{eqn:cpscov}) is derived assuming
the flat-sky approximation as we focus mainly on small angular scale
information, but the full-sky expression can be derived combining the
methods developed in this paper and in \cite{Hu00}.

\begin{figure}[th]
\centerline{\psfig{file=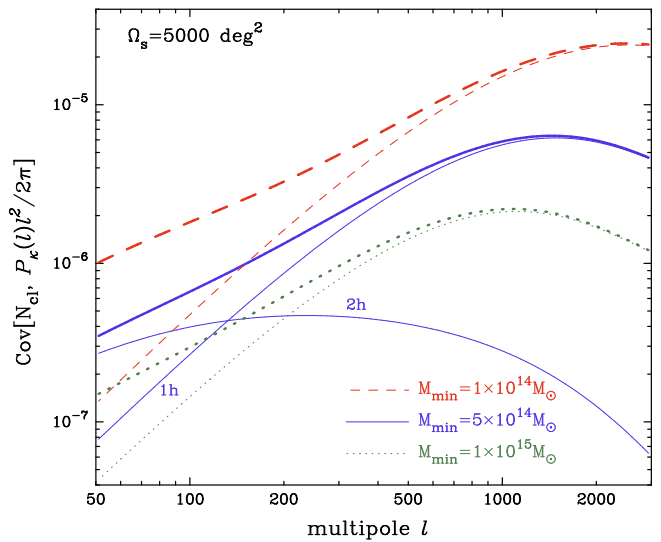, width=8.6cm}}
\caption{ The cross-covariance between the cluster
counts ${\cal N}$ and the lensing power spectrum
$P^{\rm est}_\kappa(l)$ as a
function of the angular multipole $l$ for a $\Lambda$CDM model.  Note
that
for the purpose of this Figure
we assume a single redshift bin for the lensing power spectrum for
a typical ground-based survey (see \S~\ref{survey}) and, for the cluster
counts, we include all the clusters with masses above a given minimum
halo mass over a range of redshifts $0\le z\le 1$.  The dashed, solid
and dotted curves demonstrate the results when minimum halo masses of
$M_{\rm min}=1, 5$ and $10\times 10^{14}M_\odot$ are employed,
respectively. The two thin, solid curves show the 1- and 2-halo term
contributions to the cross-covariance of $M_{\rm min}=5\times 10^{14}M_\odot$
(see Eqs.~[\ref{eqn:cpscov}] and
 [\ref{eqn:cov1h-2h}]),
while the
thin dashed and dotted curves show the 1-halo term contribution.
 } \label{fig:cov}
\end{figure}
\begin{figure}[th]
\centerline{\psfig{file=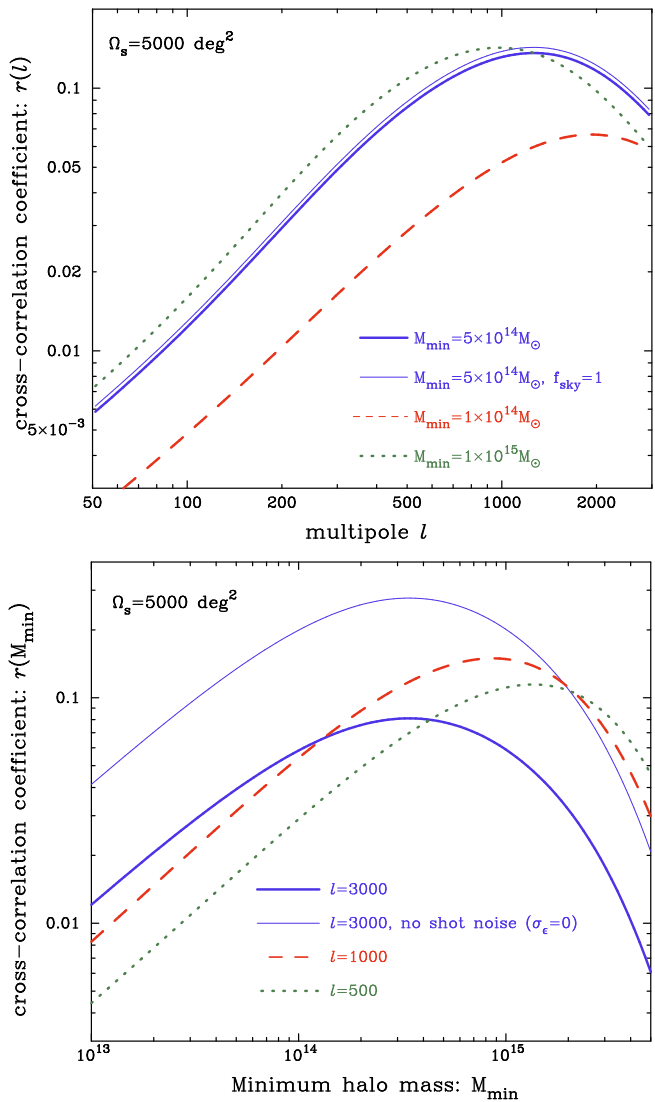, width=8.6cm}}
\caption{{\em
Upper panel}: The cross-correlation coefficients, defined by
Eq.~(\ref{eqn:r}), as a function of multipole $l$. The coefficient
depends on the multipole bin width and survey area; we assumed $\delta
l/l\simeq 0.04$ and $\Omega_{\rm s}=5000$ deg$^2$ ($f_{\rm sky}\simeq
0.12$), except for the thin solid curve where we assumed a full-sky
survey $f_{\rm sky}=1$. {\em Lower panel}: A similar plot, but as a
 function of mass thresholds in the cluster counts, for a fixed
 multipole of the lensing power spectrum. The bold solid, dashed and
 dotted curves are the results for $l=3000$, $1000$ and $500$,
 respectively. The thin solid curve shows the result for $l=3000$ if the
 intrinsic ellipticity noise is ignored.}
\label{fig:r-cov}
\end{figure}
Fig.~\ref{fig:cov} shows the cross-covariance between
the mass-selected
cluster counts and the weak lensing power spectrum as a function of
angular multipole $l$, for a concordance $\Lambda$CDM model.
For illustrative clarity
we use a single redshift bin for both of the cluster
counts and the lensing power spectrum.
The
dashed, solid and dotted curves are the results obtained when minimum
halo masses of $M_{\rm min}/10^{14}M_\odot=1, 5 $ and $ 10$ are assumed
for the cluster counts, respectively.  The two thin, solid curves show
the 1- and 2-halo term contribution to the total power (bold solid
curve)
for the  $M_{\rm min}/10^{14}M_\odot=5$ mass cut.
It is apparent that the cross-covariance at small angular scales
$l\simgt 500$ arises mainly from the 1-halo term contribution. Comparing
the dashed, solid and dotted curves clarifies that the covariance
amplitude gets greater with decreasing minimum halo mass,
as the
weak lensing and the cluster counts probe more similar density fields
in the large-scale structure as implied in Fig.~\ref{fig:cl}.

A more useful quantity is
the cross-correlation
coefficients defined as
\begin{equation}
r(l)=\frac{C^{gc}_{l1}}{\sqrt{C^c_{11}C^g_{ll}}},
\label{eqn:r}
\end{equation}
where the subscript `$1$'
denotes the first redshift bin
because for this calculation we are
putting all the clusters into a single redshift bin, for
illustration (the cluster
redshift bin index
$b=1$ for this case).
The coefficients
quantify the relative importance of the cross-covariance to the
auto-covariances
at a given $l$. The upper panel of Fig.~\ref{fig:r-cov} shows
the correlation coefficients
for model parameters assumed in Fig.~\ref{fig:cov}.
The coefficients depend on the multipole bin width
taken in the lensing power spectrum covariance calculation as well as on
a survey sky coverage; we here assumed $\Delta l/l\simeq 0.04$ and
$\Omega_{\rm s}=5000$ deg$^2$, except for the thin solid curve where a
full-sky survey $f_{\rm sky}=1$ is considered.

The upper panel of Fig.~\ref{fig:r-cov} shows
that the coefficients peak
around $l\sim 1000$, and decrease at smaller scales.
On the intermediate scales there is a significant
cross-correlation since the
1-halo term in the lensing power spectrum depends so strongly
on the number of clusters (Fig.~\ref{fig:cl}).
However on smaller scales the lensing covariance is dominated by
shot noise in the intrinsic galaxy shapes (e.g., see Fig.~1 in
\cite{Jainetal07}), which do not correlate with the cluster counts.
Comparing the thin and bold solid curves shows
that the coefficients have only weak dependence on the sky coverage,
reflecting that the sampling variance of the cluster count covariance
(\ref{eqn:clcov}) roughly scales as $f_{\rm sky}^{-1}$ for a large area
survey of our interest, which is same dependence as the other elements
in the covariances.

The lower panel of Fig.~\ref{fig:r-cov}
shows the correlation
coefficients with varying mass thresholds in the cluster counts, for a
fixed multipole $l$ of the lensing power spectrum. The lensing power
spectrum at $l\sim 1000$ is found to be most correlated with the cluster
counts for the $M_{\rm min}\sim 10^{15}M_\odot$ mass cut.
The correlation decreases at high mass thresholds when the number of
clusters is very small and therefore not representative of the
lensing field.
The correlation also decreases at smaller masses since the contribution
to the lensing power spectrum is small for light halos (Fig~\ref{fig:cl}).
As can be seen
from comparison of the bold and thin solid curves, an inclusion of
the intrinsic ellipticity noise suppresses the correlation coefficients.

\begin{figure}[ht]
\centerline{\psfig{file=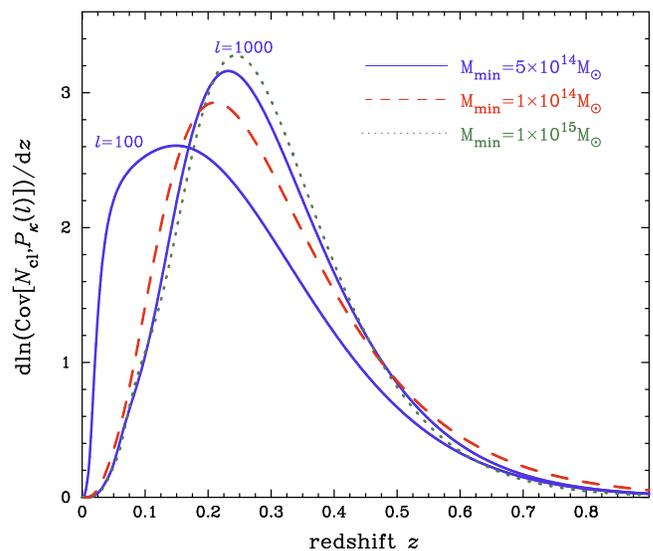, width=8.5cm}}
\caption{The relative contribution of each redshift clusters to the
cross-covariance at a given multipole.  We assumed the same model
parameters as in Fig.~\ref{fig:cov}. For an angular scale of $l=1000$,
clusters at $z\sim 0.2$ most contribute to the cross-covariance,
reflecting the peak of the lensing efficiency function.
Note that we are here using the simple mass threshold for the cluster
selection, not the lensing-based selection.
 On the other
hand, for a larger angular scale of $l=100$, most of contribution comes
from clusters at lower redshifts.}  \label{fig:cov_z}
\end{figure}
We now consider multiple redshift bins in the cluster catalog.
Fig.~\ref{fig:cov_z} shows the relative contribution of each
cluster count redshift bin
to the cross-covariance at a given multipole.
It is
clear that clusters at $z\sim 0.2-0.3$ contribute most to the
cross-covariance for an angular scale of $l\sim 1000$.
Since the number density of mass-selected cluster counts has a weak
redshift dependence as shown in Fig.~\ref{fig:dndz}, one can notice that
the redshift peak reflects redshift dependence of the lensing
efficiency function for a source redshift $z_s\sim 1$ that corresponds
to the survey depth we assumed; structures at $z\sim 0.2-0.3$ most
efficiently cause the lensing effect on source galaxies at $z_s\sim
1$. It is also worth noting that clusters at higher redshifts have a
smaller
angular
size (smaller virial radii) than
$\theta\sim 1/l$
(e.g. see the right panel
of Fig.~2 in \cite{TJ03b}).  In other words, clusters at $z\simgt 0.5$
may
carry complementary information to the lensing power spectrum.
On the other hand, for an angular scale
of $l\sim 100$, clusters at lower redshifts $z\sim 0.1$ contribute most
to the covariance, because the cluster virial radius
matches such a large angular
scale only if the cluster is located at lower redshift.

\section{Results: Signal-to-Noise and Parameter Forecasts}
\label{results}

\subsection{A CDM model and survey parameters}
\label{survey}

To compute the observables of interest
we need to specify cosmological model and
we assume survey parameters similar to those of future surveys in order to
estimate realistic measurement errors.

We include
the key parameters that may affect the observables within
an adiabatic CDM dominated model with dark energy component: the density
parameters are $\Omega_{\rm de}(=0.73)$, $\Omega_{\rm m}h^2(=0.14)$, and
$\Omega_{\rm b}h^2(=0.024)$ (note that we assume a flat universe); the
primordial power spectrum parameters are the spectral tilt, $n_s(=1)$,
the running index, $\alpha_s(=0)$, and the normalization parameter of
primordial curvature perturbation, $\delta_\zeta(=5.07\times 10^{-5})$
(the values in the parentheses denote the fiducial model). We employ the
transfer function of matter perturbations, $T(k)$, with baryon
oscillations smoothed out \cite{EisensteinHu99}.
We employ
the dark energy model \cite{ChevallierPolarski01,Linder03}
parametrized as
 $w(a)=w_0+w_a(1-a)$, with fiducial
values $w_0=-1$ and $w_a=0$.

We specify survey parameters that well resemble a future ground-based survey
(e.g., see
\cite{HSC}).
We model the redshift distribution of galaxies by using a toy
model given by Eq.~(4) in \cite{Hutereretal06}; we employ the parameter
value $z_0=0.3$ leading the redshift distribution to peak at $z_{\rm
peak}=2z_0=0.6$ and have a mean redshift of $z_{\rm m}=3z_0=0.9$. The
intrinsic ellipticities dilute the lensing shear measurements
according to Eq.~(\ref{eqn:psobs});
we simply
assume that the shot noise contamination is modeled by the rms
ellipticity per component, $\sigma_\epsilon=0.22$, and the total number
density of galaxies, $\bar{n}_g=30$ arcmin$^{-2}$.
The survey area is taken to be
 $\Omega_{\rm s}=5000 $ degree$^2$ for our
fiducial choice.
Note that throughout this paper we will assume that the two
observables we are interested in, the cluster number counts and the
lensing power spectrum, are taken from the same survey region, to study
how the cross-correlation affects the parameter constraints.
 as the two
methods probe the same cosmic structure.

\subsection{A signal-to-noise ratio}

\begin{figure}[ht]
\centerline{\psfig{file=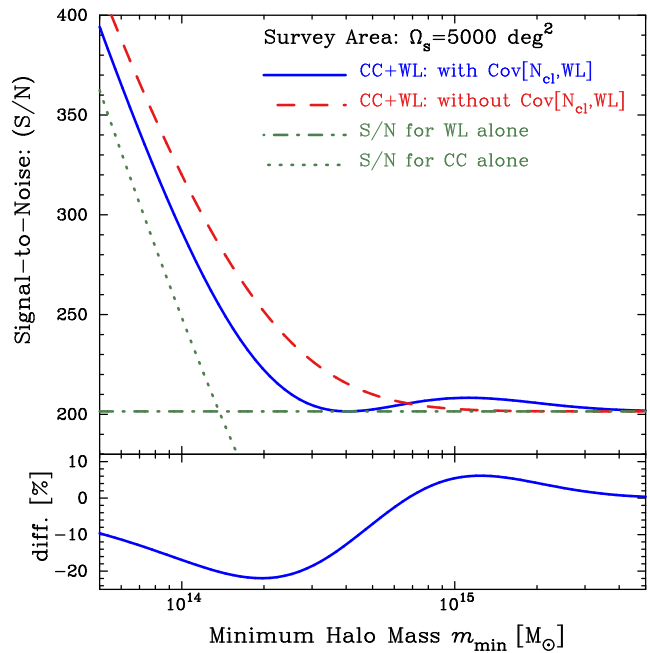, width=8.5cm}}
\caption{{\em Upper panel}:
Signal-to-noise ($S/N$) ratio as a function of the mass threshold of the cluster
count measurement.
The solid curve shows the total $S/N$ for a combined measurement of the cluster
counts and lensing power spectrum
where the cross-correlation between the two observables,
${\rm Cov}[{\cal N}_{\rm cl}, P_{\kappa}(l)]$, is included.
For comparison, the dashed curve shows the $S/N$
assuming that the two observables are independent (i.e. the
cross-correlation is ignored). The dotted and dot-dashed curves show the
$S/N$ when either of the cluster counts or the lensing power spectrum
alone is considered. {\em Lower panel}: The
percentage difference in $S/N$
with and without the cross-covariance ${\rm Cov}[{\cal
N}_{\rm cl},P_\kappa(l)]$ (i.e., the difference between the solid and
dashed curves
divided by the dashed curve
in the upper panel $\times 100$).  Interestingly, the cross-correlation
improves the $S/N$ by up to $10\%$ when only massive clusters with
$M\simgt 10^{15}M_\odot$ are included in the counts (see text for a more
detailed discussion).
All curves assume a survey with an area of $\Omega_s=5000 $ deg$^2$.
Note that
we considered a single redshift bin for both of the two observables, and
included the number counts of clusters with masses greater than the mass
threshold over redshifts $0.05\le z\le 1.0$ and the lensing power
spectrum at multipoles over $50\le l \le 3000$, respectively (see text
for the details).
} \label{fig:sn}
\end{figure}
\begin{figure}
\centerline{\psfig{file=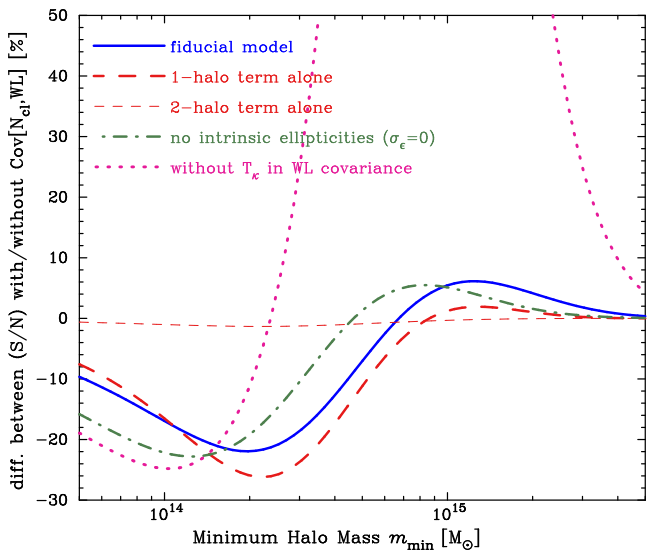, width=8.5cm}}
\caption{Shown is the relative importance of each ingredient in the
covariance calculation to the percentage difference in $S/N$. The solid
curve shows the result for our fiducial model, as shown in the lower
panel of Fig.~\ref{fig:sn}. The dashed curve shows the results where
only the 1-halo terms are included in each element of the full
covariance matrix, while the thin dashed curve shows the results when
only the 2-halo term contribution to the cross-covariance is included.
The dotted and dot-dashed curves show the results when the intrinsic
ellipticity noise or the non-Gaussian errors are ignored in the weak
lensing power spectrum covariance, respectively.  } \label{fig:sn_1halo}
\end{figure}
It is instructive to investigate the expected signal-to-noise ($S/N$)
ratio for a combined measurement of the cluster counts and the lensing
power spectrum, in order to highlight how the cross-correlation between
the two observables affects the measurement accuracies. The $S/N$ can be
estimated using the covariance matrix derived in \S~\ref{cov} as
\begin{eqnarray}
\left(\frac{S}{N}\right)^2_{c+g}\equiv
\sum_{i,j}
D_i
\left[(\bm{C}^{g+c})^{-1}\right]_{ij}
D_j.
\label{eqn:sn}
\end{eqnarray}
Here the data vector of our observables, $\bm{D}$, constructed from the
lensing power spectrum tomography with $n_s$ redshift bins and the
cluster number counts with $b$-redshift bins
is
defined as
\begin{equation}
\bm{D}=\left\{
{P_{(11)\kappa}(l_1)
,\cdots, P_{(n_sn_s)\kappa}(l_{\rm max})},
N_{(1)},\cdots,N_{(b)}
\right\}.
\label{eqn:datavector}
\end{equation}
Note that the dimension of $\bm{D}$ is $b + n_l \times n_s(n_s+1)/2$
when the lensing tomography with $n_l$ multipole bins and $n_s$ redshift
bins and the cluster counts with $b$  redshift bins are considered (also
see below Eq.~[\ref{eqn:lenstrisp}]). For a case of $b=10$, $n_s=3$ and
$n_l=100$, the dimension of $\bm{D}$ is 610.
The full covariance matrix for the
joint measurement, $\bm{C}^{c+g}$, can be constructed from
Eqs.~(\ref{eqn:clcov}), (\ref{eqn:pscov}), and (\ref{eqn:cpscov}) as
\begin{eqnarray}
\bm{C}^{g+c}\equiv
\left(
\begin{array}{cc}
\bm{C}^g & \bm{C}^{gc}\\
\bm{C}^{gc} & \bm{C}^{c}\\
\end{array}
\right).
\label{eqn:g+ccov}
\end{eqnarray}
Note that $(\bm{C}^{g+c})^{-1}$ appearing in Eq.~(\ref{eqn:sn})
is the inverse matrix of $\bm{C}^{g+c}$.

For comparison we consider the $S/N$ from each of cluster counts
and weak lensing alone by using the relevant part of the data
vector in Eq.~(\ref{eqn:datavector}) and the covariance matrices.
We also compare with the $S/N$ if the cross-correlation is not
taken into account, i.e. a matrix of zeros is used instead of
the matrix $\bm{C}^{gc}$ in Eq.~(\ref{eqn:g+ccov}).
In this case that the two are independent, e.g. measured from
non-overlapping two survey regions,
the $S/N$ values from
each of the cluster counts and the lensing power spectrum alone
therefore add in quadrature to form the joint $S/N$.

When computing $S/N$ in Eq.~(\ref{eqn:sn}) care must be taken with
numerical accuracy of the matrix inversion.
The observables of interest, the angular number
density of clusters and the lensing power spectrum, have different units
and their amplitudes could therefore differ from each other by many
orders of magnitudes. To avoid numerical inaccuracies caused by this
fact, we have used the dimension-less covariance matrix
$\tilde{\bm{C}}^{g+c}$ normalized by the data vector as
\begin{equation}
\left[\tilde{\bm{C}}^{g+c}\right]_{ij}\equiv\frac{
\left[\bm{C}^{g+c}\right]_{ij}}{D_iD_j}.
\label{eqn:g+ccov2}
\end{equation}
In terms of the re-defined covariance matrix, the total $S/N$ can be
computed simply as
$(S/N)_{g+c}^{2}=\sum_{i,j}[\tilde{\bm{C}}^{g+c}]^{-1}_{ij}$.

Fig.~\ref{fig:sn} shows the $S/N$ ratios expected
for a ground-based
survey with area $\Omega_{\rm s}=5000$ deg$^2$ and
our fiducial $\Lambda$CDM model,
as a function of minimum
halo mass used in the mass-selected cluster counts.
Here we include all the clusters
with masses greater than a given mass threshold over a range of $0.05\le
z\le 1$,
and include the lensing power spectrum
at multipoles $50\le l\le 3000$ assuming the redshift distribution of
galaxies described in \S~\ref{survey}.
Note that we here consider a single
redshift bin for both
the cluster counts and cosmic shear power spectrum for simplicity,
and the
signal-to-noise ratios only slightly increase by
adding redshift binned information (e.g., see Fig. 5 in \cite{TJ04}).
First of all, we
should notice that the lensing power spectrum and the cluster number
counts have similar $S/N$ ratios, when the mass threshold $M_{\rm
min}\sim 10^{14}M_\odot$ is used. At mass thresholds smaller than
$10^{14}M_\odot$, the cluster counts (dotted curve) have a greater $S/N$
than the lensing power spectrum (dot-dashed curve) due to an increase in
the number of sampled clusters, while the lensing power spectrum has a greater
$S/N$ at the greater mass threshold.

The solid curve shows the total $S/N$ for a combined measurement of the
cluster counts and the lensing power spectrum, when the
cross-correlation between the two observables is correctly taken into
account for the full covariance matrix (see Eq.~[\ref{eqn:g+ccov}]).
We compare this to the standard approach in which the two probes
are considered to be independent (dashed curve).
The lower panel explicitly shows the
percentage difference in $S/N$
with and without the cross-covariance.

At
small mass thresholds $\simlt 10^{14}M_\odot$, the total $S/N$
taking into account the cross-covariance
is degraded compared to
when the probes are considered independent. This is
because the
 cosmic density field probed is
 shared by
the two
observables
 and therefore
 an inclusion of the cross-covariance
reduces independence of the two observables.
However, the degradation ceases at a critical mass
scale where the total $S/N$
(including the covariance)
is equal to the $S/N$ for the lensing power spectrum
alone. In other words, the total $S/N$ is never smaller than the $S/N$
obtained from either alone of the lensing power spectrum or the cluster
counts.

Then, an intriguing result is found: the total $S/N$ is
slightly
\emph{improved}
by including the cross-covariance as the
mass threshold is increased up to $M\sim 10^{15}M_\odot$,
where the improvement  is up to $\sim 10\%$ as shown in the lower panel.
This occurs
 even though the $S/N$ ratio from the cluster counts alone
is much less than that for the lensing power spectrum alone.
The peak mass scale of the total $(S/N)$ corresponds to the mass
scale at which the correlation coefficient of the covariance
peaks as can be found in the lower
panel of Fig.~\ref{fig:r-cov}. That is, the improvement in $S/N$
could happen
when the two observables are highly correlated.
Since the cross-covariance describes how
the two observables are correlated with each other,
it appears that
a knowledge of the
number of such massive clusters with $M\simgt 10^{15}M_\odot$ for a
given survey region helps to improve the
amount of information that can be extracted from the
weak lensing measurement (also see \cite{Neyrincketal06,NeyrinckSzapudi07} for
the related discussion).
In
simpler words, if a smaller or greater number of massive clusters than
the ensemble average value was observed from a given survey region, the
observed lensing power spectrum will most likely have smaller or greater
amplitudes at $l\sim 10^3$,
respectively.

We reproduce this qualitative behavior using a simple toy model described in the
Appendix~\ref{appendix:toy-wl}, where the lensing power spectrum is
modeled to be given solely by the number of halos, ignoring the halo mass profile and the
clustering of different halos.
Based on this toy model we attribute the increase in the total $S/N$
for high cluster mass thresholds to the fact that the lensing power spectrum
amplitude is sensitive to the number of such massive clusters as
demonstrated in Fig.~\ref{fig:cl}.
The fact that the lensing power spectrum is sensitive to the number counts
weighted by the mass squared, means that it adds complementary information
to the unweighted sum from the cluster counts.

It should, however, be noted that the improvement in $S/N$ is
achievable only if the cross-covariance is {\em a priori} known from the
theoretical prediction
e.g. based on CDM structure formation scenarios.
Alternatively it could be obtained from
a measurement of the cross-correlation from the
survey region.

\begin{figure}
\centerline{\psfig{file=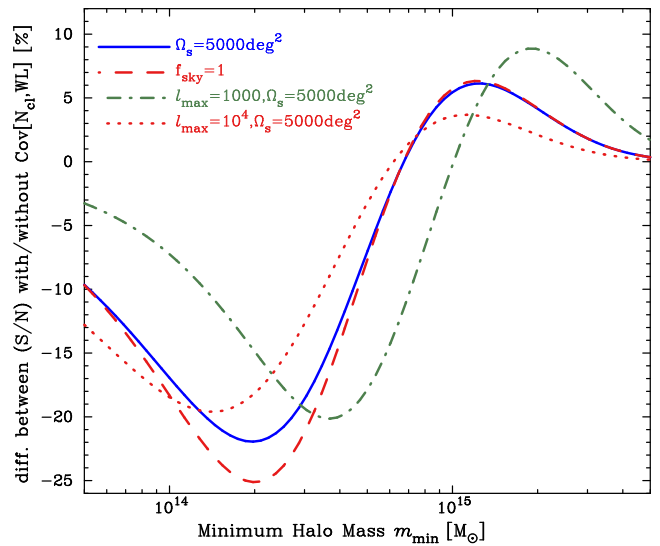, width=8.5cm}}
\caption{
Dependence of the percentage difference in $S/N$ (see lower panel of
previous plot)
on survey area and maximum multipole $l_{\rm max}$, where information on
the lensing power spectrum over a range of $50\le l\le l_{\rm max}$ is
included.
The default is $l_{\rm max}=3000$.
}
\label{fig:sn_fsky}
\end{figure}
\begin{figure}
\centerline{\psfig{file=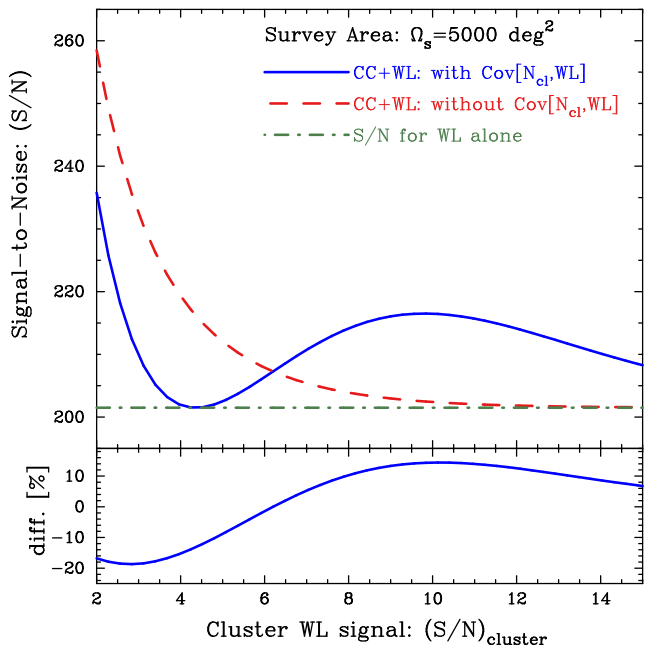, width=8.5cm}}
\caption{As in Fig.~\ref{fig:sn}, but the total signal-to-noise for the
 lensing-based cluster counts (see Fig.~\ref{fig:dndz-wl}) combined with
 the lensing power spectrum measurement is shown against detection thresholds of
 the cluster lensing signal (see Eq.~[\ref{eqn:cluster-wl}]).  Note that
 the plotting range of $y$-axis in the upper panel is smaller than that
 of Fig.~\ref{fig:sn}. } \label{fig:sn-wlsn}
\end{figure}
In Fig.~\ref{fig:sn_1halo} we study which model ingredient in the
full-covariance calculation mainly leads to
the results in
Fig.~\ref{fig:sn}. The dashed curve shows the percentage
difference in $S/N$ when only the
1-halo terms are included in each element of the full covariance matrix
(\ref{eqn:g+ccov}), which corresponds to a simplified case that there is
no clustering between halos.
Compared to
the solid line, or
Fig.~\ref{fig:sn}, the results are little different.  The dot-dashed
curve shows the result obtained when we ignore the intrinsic ellipticity
noise that contributes only to the diagonal elements of the weak lensing
power spectrum covariance. Again only a small difference is
found. On the other hand, the thin dashed curve
shows the results when only the 2-halo term contribution to the
cross-covariance is included, which
attempts to reproduce the results in the previous work \cite{FangHaiman06}.
For this case, the impact of the cross-covariance on the $S/N$ is
negligible as concluded in \cite{FangHaiman06}.
Rather, it turns out that the most
important effect comes from  the lensing trispectrum contribution to the
lensing power spectrum covariance. If we switch off the non-Gaussian contribution,
the percentage difference in $S/N$ is significantly changed. In particular,
 there is a significant improvement
in $S/N$ by adding the cluster counts with $M\simgt 10^{15}M_\odot$,
because ignoring the lensing trispectrum
decreases the diagonal elements in the full covariance
matrix and thus
enhances the relative
importance of the cross-covariance.
 This also implies that the cosmic shear fields are
highly non-Gaussian as carefully investigated in \cite{TJ07}.

Fig.~\ref{fig:sn_fsky} demonstrates how
this
percentage difference in $S/N$
depends on the sky coverage
($f_{\rm sky}$) and the maximum multipole ($l_{\rm max}$) of the lensing
power spectrum.  All the curves in Fig.~\ref{fig:sn_fsky} are very
similar, showing a weak dependence on $f_{\rm sky}$ and $l_{\rm max}$.
(Note however that the absolute
$S/N$ itself has a strong dependence.) Nevertheless, there
are several
points to note when interpreting the results.
Comparing the dotted, solid and
dot-dashed curves clarify that, with increasing $l_{\rm max}$,
the mass threshold corresponding to the dip in the percentage
difference in $S/N$
increases. This is because the lensing power spectrum at higher
multipoles is more sensitive to the cosmic density fields down to
smaller length scales, and therefore the cluster counts including less
massive halos are more correlated with the lensing power spectra.
Also
the impact of the cross-covariance is reduced when
assuming $l_{\rm max}=10^4$, compared to the fiducial case of $l_{\rm
max}=3000$, because the intrinsic ellipticity noise (shot noise) is
dominant in the lensing power spectrum covariance  at
such small scales.

Similar to Fig.~\ref{fig:sn}, Fig.~\ref{fig:sn-wlsn} shows the
total $S/N$ ratios for a combined measurement of the
\emph{lensing-based}
cluster counts and the lensing power spectrum, as a function of the
cluster lensing-signal thresholds (see Eq.~[\ref{eqn:cluster-wl}] and
Fig.~\ref{fig:dndz-wl}).  Notice that the plotting range of $y$-axis is
roughly half of that in Fig.~\ref{fig:sn}. Because the number densities
for the lensing signal thresholds of interest are less than that for the
mass-selected cluster sample (as shown in Fig.~\ref{fig:dndz-wl}) the
lensing-based cluster counts do not contribute much to the total $S/N$
ratios, compared to Fig.~\ref{fig:sn}.  Other than this difference, the
behavior for the $S/N$ curves found in Fig.~\ref{fig:sn} are similar
to those in Fig.~\ref{fig:sn-wlsn}.

\subsection{Fisher analysis for cosmological parameter constraints}
\label{fisher}

We now estimate accuracies of cosmological parameter
determination, given the measurement accuracies of the observables,
using the Fisher matrix formalism \cite{Fisher35,Tegmarketal97}.
This formalism assesses how well given observables can constrain
cosmological parameters around a fiducial cosmological model.
The parameter
forecasts we obtain depend on the fiducial model and are also sensitive
to the choice of free parameters.
Furthermore, the Fisher matrix gives only a lower limit to the
parameter uncertainties, being exact if the likelihood surface
around the local minimum is Gaussian in multi-dimensional parameter space.
Ideally
a more
quantitative method
would be used
to explore the global structure
to realize
more accurate parameter forecasts.
As described in
\S~\ref{survey}, we include all the key parameters that can describe
varieties in the observables within $\Lambda$CDM model cosmologies.

The Fisher information matrix available from the lensing power spectrum
tomography is given by
\begin{eqnarray}
[\bm{F}^{g}]_{\alpha\beta}
=\sum_{m,n}\frac{\partial P_{(ij)\kappa}(l_a)}{\partial p_\alpha}
\left[\bm{C}^g\right]^{-1}_{mn}
\frac{\partial P_{(ij)\kappa}(l_b)}{\partial p_\alpha},
\end{eqnarray}
where the partial derivative with respect to the $\alpha$-th
cosmological parameter $p_\alpha$ is evaluated around the fiducial
model, with other parameters $p_\beta$ ($\alpha\ne\beta$) being fixed to
their fiducial values.
The error on a parameter $p_\alpha$, marginalized over other
parameter uncertainties, is given by
$\sigma^2(p_\alpha)=(\bm{F}^{-1})_{\alpha\alpha}$, where $\bm{F}^{-1}$
is the inverse of the Fisher matrix.

Similarly, the Fisher matrix for the cluster counts is given by
\begin{eqnarray}
[\bm{F}^{c}]_{\alpha\beta}
=\sum_{b,b'}\frac{\partial N_{(b)}}{\partial p_\alpha}
\left[\bm{C}^c\right]^{-1}_{bb'}
\frac{\partial N_{(b')}}{\partial p_\alpha}.
\end{eqnarray}

For a combined measurement of the lensing power spectrum and the
cluster counts, the Fisher matrix is calculated using the full
covariance matrix defined by Eq.~(\ref{eqn:g+ccov}) (also see
Eq.~[\ref{eqn:g+ccov2}]) as
\begin{eqnarray}
[\bm{F}^{g+c}]_{\alpha\beta}
=\sum_{i,j}
\frac{\partial \ln D_i}{\partial p_\alpha}
\left[
(\tilde{\bm{C}}^{g+c})^{-1}\right]_{ij}
\frac{\partial \ln D_j}{\partial p_\alpha},
\end{eqnarray}
where the summation indices $i,j$ run over the redshift
and
multipole bins of the tomographic lensing power spectra as well as the
redshift bins of the cluster counts.

Using probes of structure formation alone
is not powerful enough to
constrain all the cosmological parameters simultaneously and well.  Rather,
combining the large-scale structure probes with constraints from
CMB temperature and polarization anisotropies significantly helps
to lift parameter degeneracies and, in particular, the dark energy
parameters (e.g. \cite{Eisensteinetal99,TJ04}).
When
computing the Fisher matrix for a given CMB data set, we employ 9
parameters: the 8 parameters described in \S~\ref{survey} plus the
Thomson scattering optical depth to the last scattering surface,
$\tau(=0.10)$.  Note that we ignore the contribution to the CMB spectra
from the primordial gravitational waves for simplicity.  We use the
publicly-available CMBFAST code \cite{SeljakZaldarriaga96} to compute
the angular power spectra of temperature anisotropy, $C^{\rm TT}_l$,
$E$-mode polarization, $C^{\rm EE}_l$, and their cross-correlation,
$C^{\rm TE}_l$.  To model the measurement accuracies we assume the noise
per pixel and the angular resolution of the Planck experiment that were
assumed in \cite{Eisensteinetal99}.

To be conservative, however, we do not include the CMB information on
the dark energy equation of state parameters, $w_0$ and $w_a$.  We do
this because essentially angular positions of the CMB acoustic peaks
constrain a degenerate
combination of the curvature of the universe and the dark energy
parameters, through their dependences on  the angular diameter
distance to the last scattering surface.
We are assuming a flat universe and
therefore wish to remove the artificially good constraint on dark energy
that we would get from the CMB.
Note, however, that our parameter forecasts shown below would not
change
significantly 
even for a non-flat universe, because we focus on
large-scale structure probes
{\em in combination with } the CMB constraints, as carefully shown in
\cite{Knoxetal06}.

We remove the CMB information on the dark energy parameters
using the following steps.
We first compute the inverse of the CMB Fisher matrix, $\bm{F}^{-1}_{\rm
CMB}$, for the 9 parameters in order to obtain marginalized errors on
the parameters, and then re-invert a sub-matrix of the inverse Fisher
matrix that includes only the rows and columns for the parameters
besides $w_0$ and $w_a$. The sub-matrix of the CMB Fisher matrix derived
in this way describes accuracies of the 7 parameter determination,
having marginalized over degeneracies of the dark energy parameters
$w_0$ and $w_0$ with other parameters for the hypothetical Planck data
sets.
In addition, we use only the CMB information in the range of
multipoles $10\le l\le 2000$, and therefore we do not include the
integrated Sachs-Wolfe (ISW)
effect that contribute to the CMB spectra
mainly at low multipoles $l\simlt 10$,
because the ISW effect is very likely
correlated with the cosmic shear power spectrum and cluster counts, and
we will ignore the correlations in this paper.

To obtain the Fisher matrix for a joint experiment combining the lensing
and/or cluster experiments with the CMB information, we simply sum the
two Fisher matrices as, e.g. $\bm{F}^{g+c+{\rm CMB}}=\bm{F}^{g+c}
+\bm{F}^{\rm CMB}$ because the CMB information can be safely considered
as an independent probe to the low-$z$ universe probes in our setting.
Note that the
final Fisher matrix such as $\bm{F}^{g+c+\rm CMB}$ has $9\times 9$ dimensions.

\subsection{Forecasts for parameter constraints}
\begin{figure*}[t]
\centerline{\psfig{file=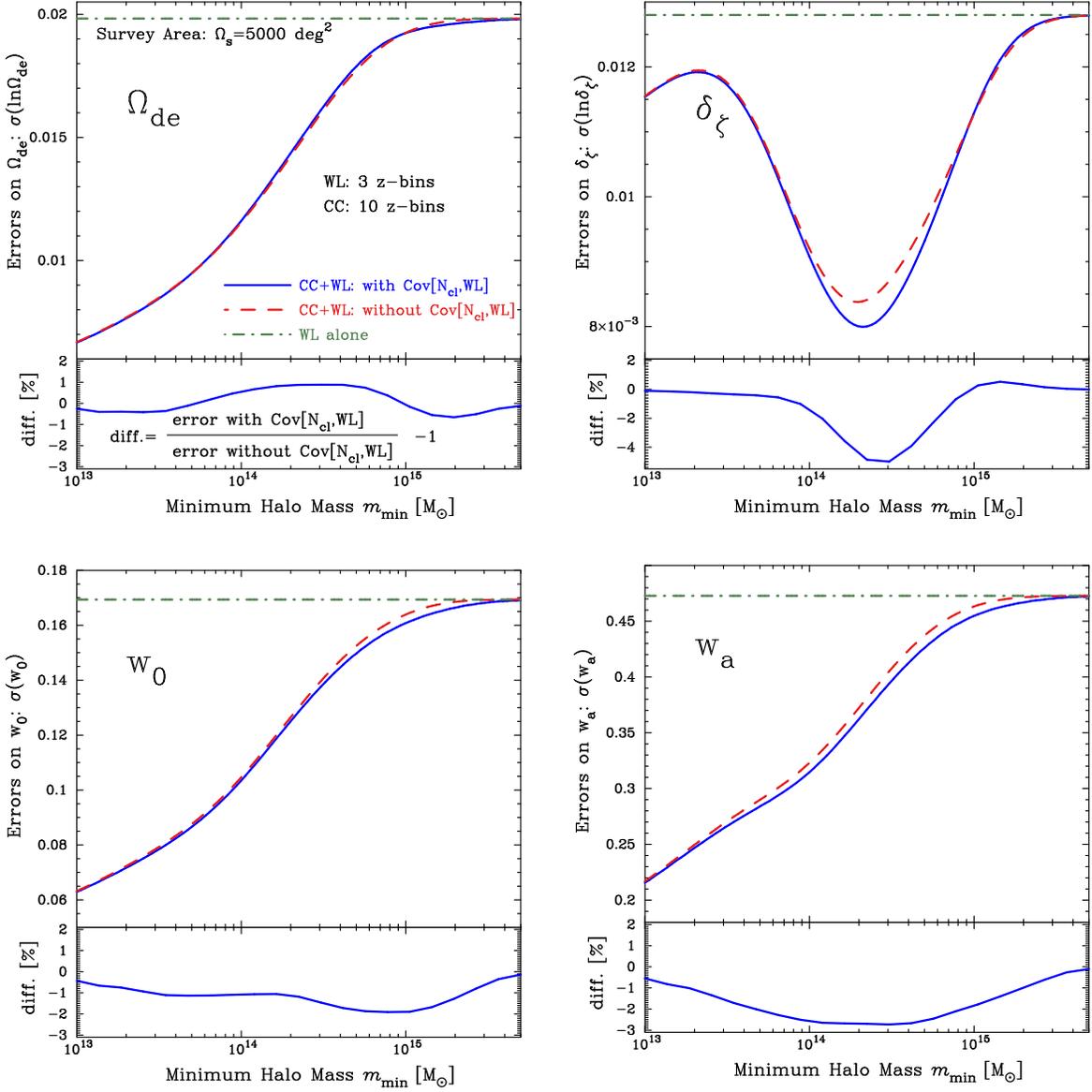, width=15.5cm}}
\caption{The projected $68\%$ C.L. error on each of the parameters,
$\Omega_{\rm de}$ (upper-left panel), $\delta_\zeta$ (upper-right),
$w_0$ (lower-left) and $w_a$ (lower-right), marginalized over other
parameters (9 parameters in total), as a function of mass thresholds
used in the cluster counts.  We assume 10 redshift bins for the cluster
counts over redshifts $0.05\le z\le 1.0$, and 3 redshift bins for the
lensing power spectrum tomography assuming the redshift distribution of
galaxies described in \S~\ref{survey}, for a survey of 5000 deg$^2$
area.  In the {\em upper panel} of each plot, the solid curve shows the
errors expected from a combined measurement of the cluster counts and
the lensing tomography when the cross-covariance between the two
observables are correctly taken into account,
while the dot-dashed curves shows the errors for the lensing
tomography alone.
The dashed curve shows the error
from combining cluster counts and lensing tomography
when the cross-covariance is ignored. Note that all the results shown
here assume the Planck priors discussed in \S~\ref{fisher}. The {\em
lower panel} of each plot shows the percentage difference in the
parameter errors with and without the cross-covariance, highlighting the
impact of the cross-covariance on the parameter forecasts.  The errors
on the dark energy equation of state parameters, $w_0$ and $w_a$, are
improved by an inclusion of the cross-covariance over mass thresholds we
 have considered, but the effect is small (only a few \%).}
\label{fig:errors}
\end{figure*}
\begin{figure*}[t]
\centerline{\psfig{file=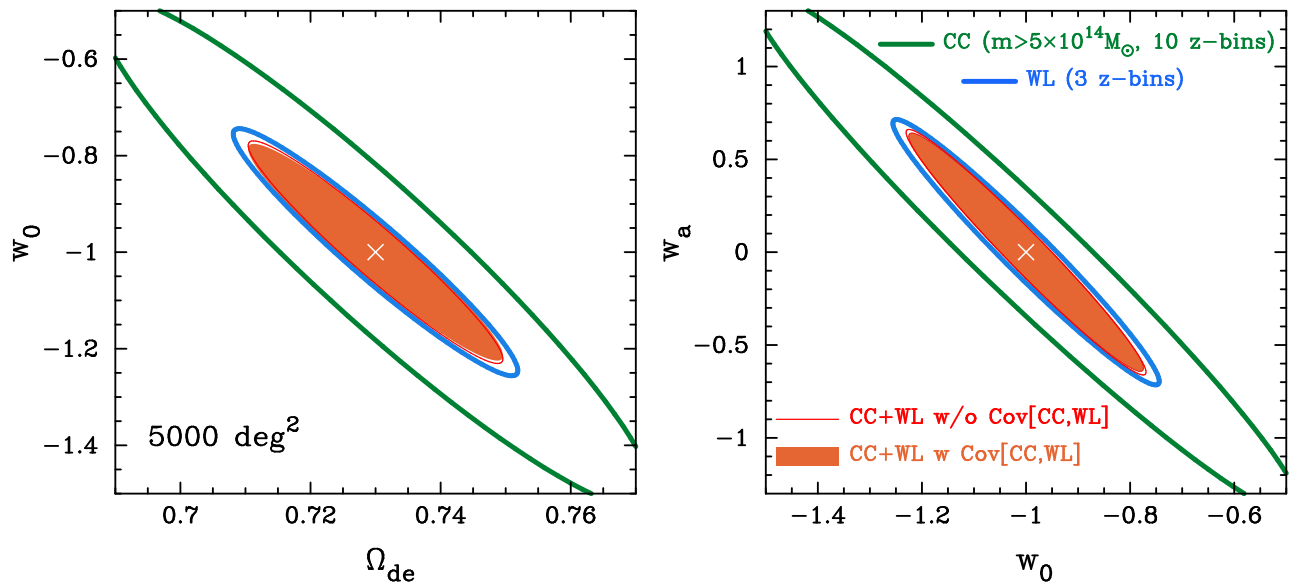, width=17.cm}}
\caption{As in Fig.~\ref{fig:errors}, but this plot explicitly shows
projected 68\% error ellipses in two-parameter subspace the dark energy
parameters $(\Omega_{\rm de}, w_0, w_a)$, for cluster counts of mass cut
$M_{\rm min}=5\times 10^{14}M_\odot$.  The outermost and middle
bold-solid curves in each panel are the error ellipses expected when
using either alone of the cluster counts or the lensing power spectrum
tomography in combination with Planck, while the innermost shaded
ellipses show the errors for the two observables combined. For
comparison, the thin-solid curves show the error ellipses obtained when
ignoring the cross-covariance; the effect is very small (the area is
enlarged only by a few $\%$).  The cross symbol in each plot shows the
fiducial model for the Fisher matrix analysis.}
\label{fig:halom-ellipse}
\end{figure*}
\begin{figure*}[t]
\centerline{\psfig{file=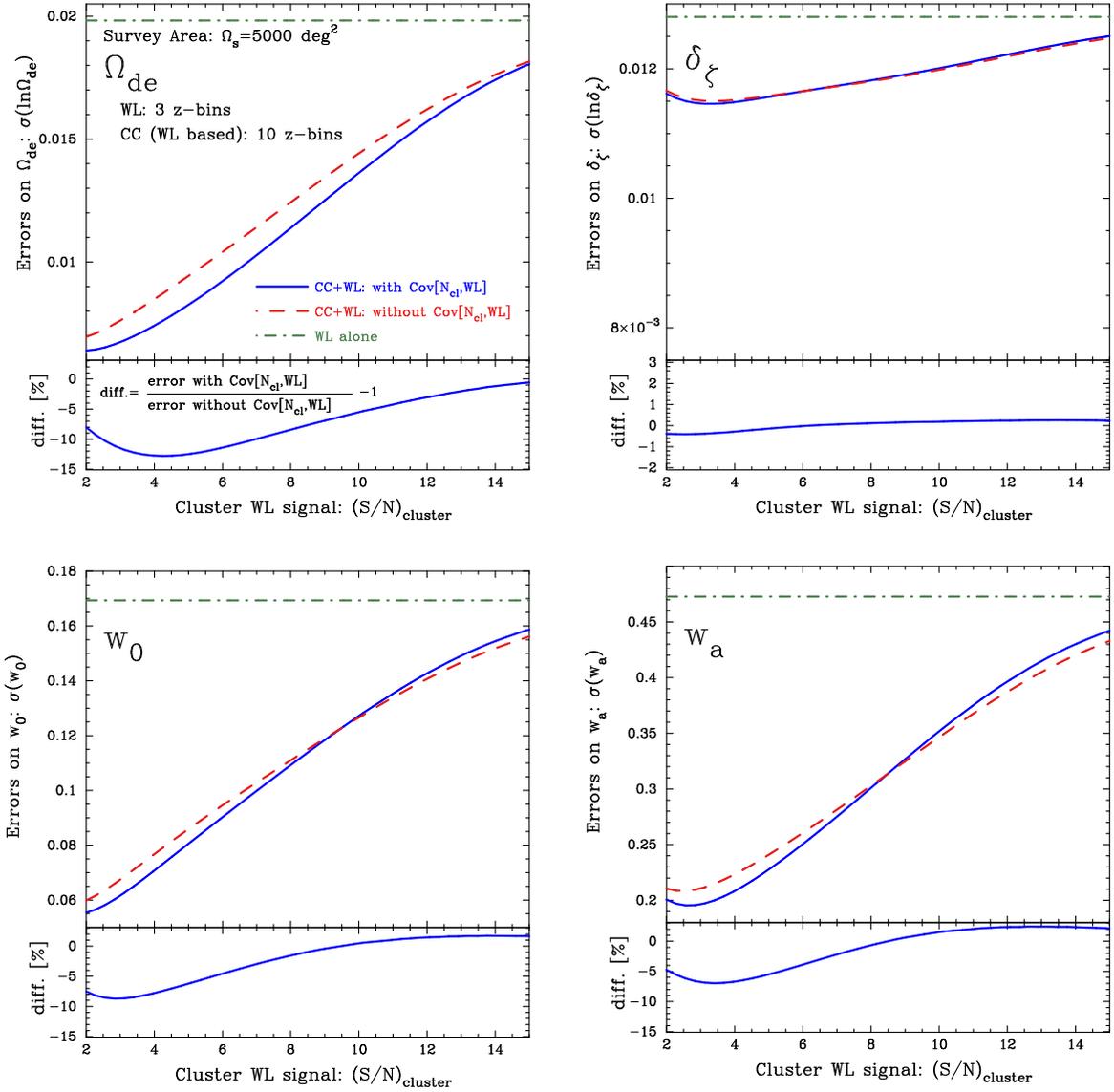, width=15.5cm}}
\caption{Similar to the previous plot, but for the lensing-based
 cluster counts as a function of the detection thresholds,
 where clusters having the lensing signal greater than the threshold are
 included in the counts. Note that the plotting range of $y$-axis in the
 upper panel of each plot is same as that in Fig.~\ref{fig:errors}.  A
similar
improvement in the parameter error is obtained by adding
 the cluster counts,
even
 though the lensing based cluster counts generally contain fewer
 clusters than the mass-selected cluster counts as shown in
 Fig.~\ref{fig:dndz-wl}.
 The inclusion of the covariance between cluster counts and lensing
 tomography is slightly larger here, compared to when a mass threshold is used
 for the cluster counts.
} \label{fig:errors-wlsn}
\end{figure*}
\begin{figure*}[t]
\centerline{\psfig{file=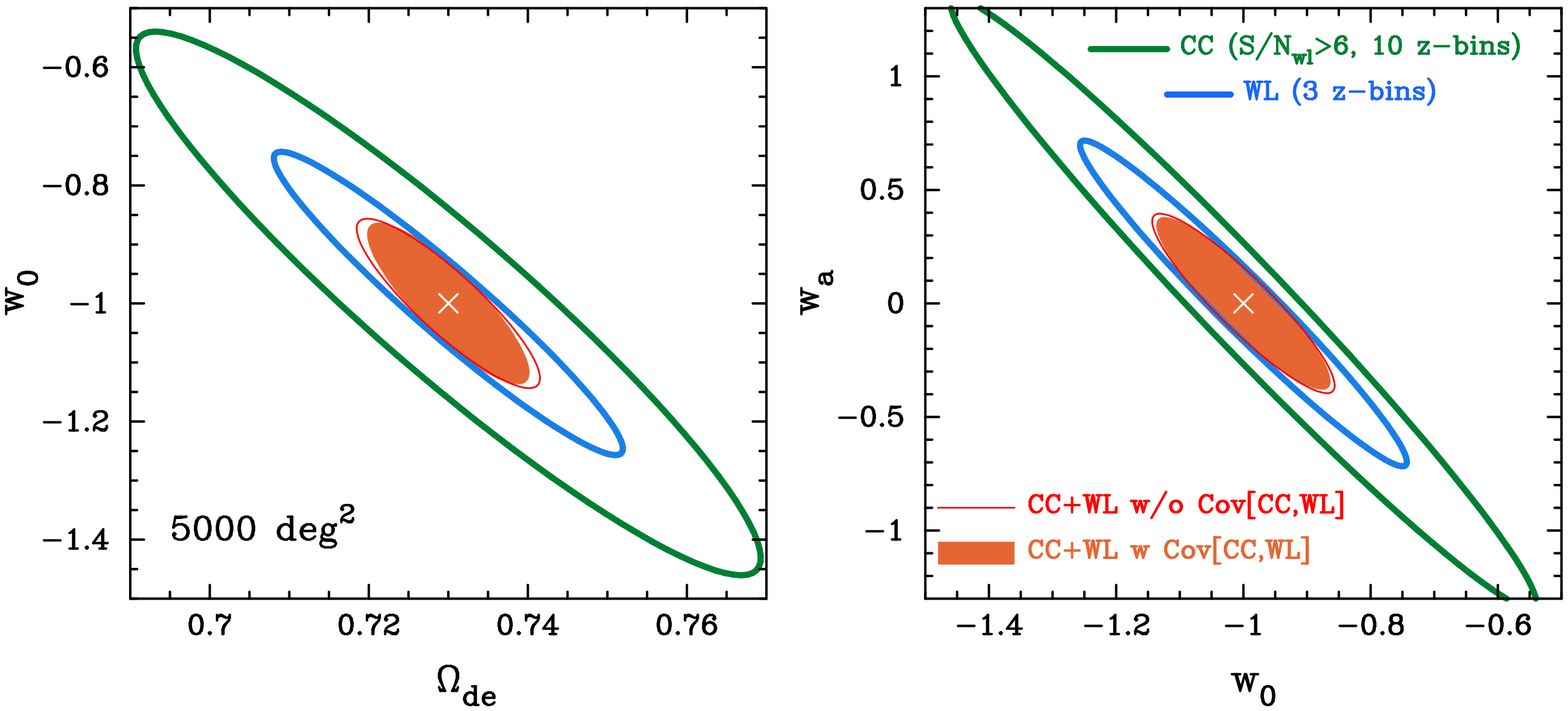, width=17.cm}}
\caption{As in Fig.~\ref{fig:halom-ellipse}, but for the lensing-based
cluster counts of detection threshold $(S/N)_{\rm cluster}=6$.}
\label{fig:wlsn-ellipse}
\end{figure*}

When forecasting  cosmological parameter determination, we should
notice that
the redshift information inherent in the cluster number counts and the
lensing power spectrum can be very powerful to significantly tighten
cosmological parameter errors, especially dark energy parameters (e.g.,
see \cite{TJ04}). In the following,
we will
assume $3$ redshift bins for lensing tomography and $10$ redshift bins
for the cluster number counts over $0.05\le z \le 1.0$, for a survey
with 5000 deg$^2$ area.
The `three' redshift bins for
lensing tomography is a minimal choice to obtain non-degenerate
constraints on the `three' dark energy parameters, $\Omega_{\rm de}$,
$w_0$ and $w_a$ as implied from Fig.~3 in \cite{Maetal05}, although four
or more redshift bins lead to further improvement, albeit not so much, in
the parameter errors. Since we ignore various systematic errors for both the
cluster counts and the lensing tomography, we adopt a rather
conservative setting for the lensing tomographic binning. However note that we
have checked a different redshift binning for cluster counts in
combination with the lensing tomography does not 
change the main
results below significantly. 
An investigation into survey optimization for survey parameters
(area and depth) and redshift binning will be presented elsewhere in a
more practical manner taking into account possible effects of the systematic errors.

We first consider mass-selected
cluster counts, and Fig.~\ref{fig:errors} shows the expected $68\%$
limits 
on each of the parameters $\Omega_{\rm de}$, $\delta_\zeta$,
$w_0$ and $w_a$,
as a function of
mass thresholds in the cluster counts.
In each case we have marginalized over the remaining 8
cosmological parameters (see \S~\ref{survey} and
~\ref{fisher} for the cosmological parameters used).
It should be also noted that the errors on these 4 parameters
are
enlarged only by $\sim 10\%$
without the CMB priors.
In the upper panel of each plot, the solid curve shows the error on a
given parameter when the cross-covariance between the two observables
is correctly taken into account,
while
the dot-dashed curve shows the error obtained from the lensing
tomography alone.  Comparing the solid and dot-dashed curves
demonstrates that adding the cluster counts for the smaller mass cuts
into the lensing tomography can
more improve the errors on dark energy parameters, $\Omega_{\rm de}$,
$w_0$ and $w_a$, because
the two observables depend on cosmological parameters in different ways
and combining the two can lift the parameter degeneracies
(also see \cite{FangHaiman06}).
To be more explicit, the errors are improved by $\sim 40\%$ for mass
threshold $M_{\rm min}\sim 10^{14}M_\odot$, while the errors are
improved only slightly by $\sim 5\%$ for $M_{\rm min}\sim
10^{15}M_\odot$.

On the other hand, there is
a complex behavior in the error on
the primordial curvature perturbation, $\delta_\zeta$.
This is explained as follows.
The cluster counts are very sensitive to the
normalization parameter of the linear mass power spectrum ($\delta_\zeta$ for
our case or $\sigma_8$ often used in the literature) through the sharp
exponential cut-off of the halo mass function at high mass end.
As we reduce the minimum  mass threshold down to
the range $3\times 10^{13}\simlt M_{\rm min}/M_\odot\simlt 10^{14}$
we are beginning to lose information about the number of very high mass
clusters, which are diluted in the total count by the large number of
low mass clusters.  At much lower mass cuts $M_{\rm
min}\simlt 3\times 10^{13}M_\odot$, the cluster counts come back to
yield a tighter constraint on $\delta_\zeta$ than the lensing tomography
through the better measurement accuracy due to the larger
number of very low mass halos.
It would be also worth pointing out that knowing the number of clusters can
effectively allow the lensing power spectrum to yield more information
on the linear theory part of the power spectrum (e.g. one could subtract
off the contribution from the clusters to get at the two halo
term). Therefore the constraint on the power spectrum amplitude can be
partly improved from the joint constraint from the mass function and the
linear power spectrum.
Note that in this paper we consider a simple mass threshold for the
cluster counts, however if clusters can be binned by mass then the
information would be restored and we may also use the shape of the
mass function to constrain cosmological parameters (e.g. \cite{Rozoetal07}).

The impact of the cross-covariance between the two observables on the
parameter forecasts can be found from comparison of the solid and dashed
curves: the dashed curve shows the error obtained when the two
observables are considered to be independent.  Further, the lower panel
of each plot explicitly presents the percentage difference in the
errors.
The impact of
the cross-covariance on the parameter errors is generally very small,
at only a few per cent.

Nevertheless, interestingly, in some cases
an inclusion of the cross-covariance leads to
an {\em improvement} in the parameter errors; for
example, the errors on $w_0$ or $w_a$ are improved over a range of the
mass thresholds we have considered.  This perhaps counter-intuitive result is
in part due to working in 9 dimensional parameter space, and
could also be explained as follows (also see \cite{TJ07} for related
discussion).  As we have carefully investigated, the cross-covariance
predicted from a CDM model quantifies how the cluster counts and the
lensing power spectrum amplitude are correlated with each other in
redshift and multipole space. The {\em positive} cross-correlation shown
in Fig.~\ref{fig:cov} implies that for a given survey region,
if the number of clusters probed
happens to be higher or lower  than the
ensemble average, the lensing power spectrum will be expected
to have  larger or smaller
amplitudes,
respectively. Therefore, such a correlated offset in the two observables
makes it difficult to determine their true amplitudes compared to the
case in which the two observables are independent, thereby degrading the
errors of parameters that are primarily sensitive to the amplitudes of
the two observables. This explains the degradation in the errors on
$\Omega_{\rm de}$ and $\delta_\zeta$ for some range of mass thresholds.
On the other hand, the correlated offset rather preserves {\em
relative} values between the cluster counts and the lensing
spectrum amplitudes in redshift and multipole space.
That is, a priori
knowledge on the cross-covariance leads to an improvement in the errors
on parameters that imprint characteristic redshift and multipole
dependences onto the cluster counts and the lensing power spectrum. This
is the case for the parameters $w_0$ and $w_a$.

In Fig.~\ref{fig:halom-ellipse} we show the projected 68\% C.L. error
ellipses in the dark energy parameter space, for one particular example
mass threshold, $M_{\rm min}=5\times 10^{14}M_\odot$. The error ellipses
in a two-parameter subspace highlight how the two parameters considered
are degenerate for a given observable and the degeneracies are broken by
combining different observables.  It can be seen that the lensing power
spectrum tomography and the cluster counts have similar degeneracy
directions in constraining the dark energy parameters.  
Adding the cluster counts only slightly improves the parameter errors
compared to the errors from the lensing tomography alone. The plot also
shows that the cross-covariance has a negligible effect on the error
ellipses.

Fig.~\ref{fig:errors-wlsn} shows the results for the lensing-based
cluster counts, as a function of the cluster lensing signal, where
clusters having a lensing signal greater than a given threshold,
$(S/N)_{\rm cluster}$, are included in the counts.  As in
Fig.~\ref{fig:errors}, we consider $10$ redshift bins for the
cluster counts over redshifts $0.05\le z\le 1$ and $3 $ redshift bins
for the lensing tomography. Note that the plotting range of $y$-axis in
the upper panel of each plot is same as that in Fig.~\ref{fig:errors},
while the plotting range in the lower panel is different.

First of all, it should
be noted that adding the lensing-based cluster counts into the
lensing tomography does tighten the errors on $\Omega_{\rm de}$, $w_0$
and $w_a$
significantly
even though the cluster counts include fewer
clusters than the mass-selected counts (see Fig.~\ref{fig:dndz-wl}). For
the threshold $(S/N)_{\rm cluster} \sim 10$, which includes clusters with
masses $M\simgt 10^{15}M_\odot$ and mainly covers a narrow redshift
range of
$0.05\simlt z\simlt 0.6$, the cluster counts still improve the dark
energy parameters by $\sim 25\%$, in contrast with only $\sim 4\%$
improvement for the mass-selected cluster counts with $M\simgt
10^{15}M_\odot$ in Fig.~\ref{fig:errors}.
We find the same percentage improvement when the CMB priors are
not included.
With the reasonable value of $(S/N)_{\rm cluster} \sim 6$
the uncertainties are halved by adding cluster counts to
lensing power spectra.
The
relatively
amplified
sensitivity to dark energy is attributed to the fact that the cluster
lensing signal itself depends on the dark energy parameters via the
lensing efficiency, even for a fixed halo mass (see
Eq.~[\ref{eqn:cluster-wl}]).

For the primordial curvature perturbation $\delta_\zeta$, adding the
cluster counts does not improve the error much, compared to the result
in Fig.~\ref{fig:errors}. The parameter $\delta_\zeta$ does not affect
the
amount of lensing for a given cluster
so the poorer accuracy just arises from larger statistical errors
due to the
smaller number of clusters,
compared to the mass-selected counts.

As shown in the lower panel of each plot, the cross-covariance has more
influence on the parameter forecasts, compared to
Fig.~\ref{fig:errors}. This is because lensing-based cluster counts
and lensing tomography both pick up halos over a very similar
range in redshifts.
Therefore
there are more significant cross-correlations between the two
observables. However,
the effect
of the cross-covariances on the parameter errors is small,
by less than $\sim 10\%$, for $(S/N)_{\rm cluster}\simgt 6$.

Fig.~\ref{fig:wlsn-ellipse} shows the marginalized error ellipses, for the
lensing-signal detection threshold
$(S/N)_{\rm cluster }=6$.  The degeneracy directions in dark energy parameter
constraints are very similar between the cluster counts and the lensing
tomography.  Compared to Fig.~\ref{fig:halom-ellipse}, the lensing-based
cluster counts have a better accuracy of constraining the dark energy
parameters, thereby leading the error ellipses to more shrink when
the cluster counts and the lensing tomography combined.
This can be explained because the contours in the dark energy 
equation of state parameter space is just a projection of 
the full 9d space. We have investigated eigendirections 
in cosmological parameter space and identified differences
between cluster counts and lensing for 
$\delta_\zeta$ and $\Omega_m h^2$ (effectively $h$ given that
$\Omega_m$ is a parameter in the Fisher matrix).
For example, we find that if we plot $w_0$ against $\delta_\zeta$ then 
we see that the cluster counts plus CMB contours are more 
aligned with the $w_0$ axis whereas the lensing plus CMB 
contours are more aligned with the $\delta_\zeta$ axis. 
When combined together, both degeneracies are broken and 
the error bar on $w_0$ is reduced. 
Similarly for $\delta_\zeta$ and $w_a$ and $\Omega_m h^2$ 
with each dark energy parameter.
This explains why the combined contours (cluster counts plus
lensing plus CMB) are smaller than either separate contour
(cluster counts plus CMB or lensing plus CMB),
even though the separate constraints have the same 
degeneracy directions when projected down onto $w_0$ versus $w_a$ 
space.

\begin{table*}
\begin{center}
\begin{tabular}{l|l@{\hspace{2em}}l@{\hspace{2em}}l@{\hspace{2em}}
l@{\hspace{2em}}|l}
\hline\hline
Probes & $\sigma(\Omega_{\rm de})$ &$\sigma(\ln\delta_\zeta)$
&$\sigma(w_{\rm piv})$&$\sigma(w_a)$&$\sigma(w_{\rm piv})\times\sigma(w_a)$
\\
\hline
WLT & 
0.014 & 0.013 & 0.039 & 0.47 & 0.019\\ \hline
CCM ($M_{\rm min}=5\times 10^{14}M_\odot$) 
& 0.028
& 0.015 & 0.085 & 0.95 & 0.081\\
WLT+CCM 
&[0.013]& [0.0096]
&[0.033]& [0.44]& [0.0143]
\\
WLT+CCM (with cross-cov.) 
&0.013 ($0\%$)& 0.0093 ($3\%$)
&0.032 ($3\%$)&
		 0.42 ($3\%$)& 0.0135($6\%$)
\\ \hline
CCWL ($(S/N)_{\rm cluster}\ge 6$)
& 0.026 & 0.015& 0.061 & 0.89 &0.054
\\
WLT+CCWL  
& [0.0076] & [0.012]&[0.038] & [0.26] &[0.00993]
\\
WLT+CCWL (with cross-cov.) 
& 0.0067 ($12\%$)&0.012 ($0\%$)& 0.038 ($0\%$)
&0.25 ($4\%$)&0.00958($4\%$)
\\
\hline\hline
\end{tabular}
\end{center}
\caption{Expected marginalized errors (68\% C.L.) for weak lensing
 tomography (WLT), the mass-selected cluster counts of mass threshold
 $M_{\rm min}=5\times 10^{15}M_\odot$ (CCM) and the lensing-based
 cluster counts of detection threshold $(S/N)_{\rm cluster}=6$ (CCWL),
 where all the probes are combined with Planck.  Here the error
 $\sigma(w_{\rm piv})$ shows the error in the equation of state at the
 best constrained redshift for a given observable.  The row labeled as,
 e.g. `WLT+CCM', shows the parameter errors expected when combining,
 e.g. `WLT' and `CCM'. The numerical values in brackets show the errors
 obtained when ignoring the cross-covariance between the cluster counts
 and the lensing tomography, while the percentage in parenthesis
 indicates improvement in the errors by including the cross-covariance.}
 \label{tab:errors}
\end{table*}
Table~\ref{tab:errors} summarizes the results shown in
Figs.~\ref{fig:halom-ellipse} and \ref{fig:wlsn-ellipse} showing the
marginalized uncertainties for determination of the 4 parameters
$\Omega_{\rm de}, \delta_\zeta, w_{\rm piv}$ and $w_a$, where $M_{\rm
min}=5\times 10^{14}M_\odot$ and $(S/N)_{\rm cluster}=6$ are employed
for the mass-selected cluster counts and the lensing-based cluster
counts, respectively. 
 The error on $w_{\rm piv}$, $\sigma(w_{\rm
piv})$, shows the error in the dark energy equation of state at the best
constrained redshift for given observables, or equivalently the error on the
constant equation of state parameter $w_0$ obtained by fixing $w_a$.
Note that the pivot redshift $z_{\rm pivot}$ is
similar for all the cases: $z_{\rm pivot}\sim 0.5$.  The numerical value
$\sigma(w_{\rm piv})\times\sigma(w_a)$ is proportional to the area of
error ellipses in the right panels of Figs.~\ref{fig:halom-ellipse} and
\ref{fig:wlsn-ellipse}. 
For the mass selected clusters there is a mild improvement
in both the error on $w_{\rm piv}$ and $w_a$ when adding
cluster counts to lensing tomography.
For the lensing selected clusters case, the improvement
is mostly in $w_a$.

\subsection{Discussion of systematic errors}
\label{sys}

We have 
considered idealized cases: we have ignored possible
systematic errors involved in both the cluster counts and the lensing power
spectrum measurement for simplicity.
In this subsection, we present some discussion of possible effects of
the ignored systematic errors on our results.

An imperfect knowledge of galaxy redshifts inferred from multi-band
imaging data
(photometric redshifts 
hereafter simply
photo-$z$) could affect both measurements of cosmic shear and cluster
counts. For cosmic shear, statistical errors in photo-$z$s are unlikely
to be a dominant source of the error budget of cosmic shear measurement,
if they are well characterized
\cite{Hutereretal06,Maetal05}. 
This is 
because gravitational lensing has a
broad redshift sensitivity function.
As carefully investigated in
\cite{Hutereretal06,Maetal05}, the dominant source of the systematic
error is rather caused by mean bias in photo-$z$s, causing mean
redshifts of the tomographic bins to be shifted relative to the true
mean redshifts. For planned future surveys, the mean redshifts need to
be known to better than a few tenths of a percent accuracy in redshift in order
to avoid much degradation in cosmological parameter errors. To achieve
this requirement, a large representative spectroscopic redshift
sub-sample of the full set of galaxies used for lensing may be needed to
calibrate/correct photo-$z$ errors.

For cluster counts, photo-$z$ errors cause uncertainties in
redshift estimates of individual clusters
and in addition
cause uncertainties in the lensing signal of individual clusters, if
a lensing-based cluster catalog is used. For the lensing signal,
the requirements on photo-$z$ accuracy would be similar to
that for cosmic shear, as discussed in the previous paragraph. To estimate
the redshift of a cluster using photo-$z$s, we often have old
red-sequence galaxies for which good photo-$z$s are easier to
obtain due to a strong 4,000$\AA$ break (e.g., \cite{gladdersy05}).
Further, the redshift of the cluster is an average over the redshifts of
the cluster members, thus reducing the uncertainty yet further.
In addition, since clusters are relatively rare objects it would not be
very expensive to perform follow-up spectroscopy on a central bright
galaxy or some member galaxies. These high-quality redshifts would allow
much finer redshift binning of the cluster distribution than redshift
bins of the cosmic shear tomography. Then, taking the cross-correlation
between the clusters with {\em known} redshifts and a fair sub-sample of
the galaxies used for the cosmic shear tomography may be used to
calibrate the photo-$z$ errors, because the cross-correlation is
non-vanishing only if the source galaxies are physically associated with
the clusters (see \cite{Zhan} for the related discussion).  
This issue would be worth exploring further, and will be
presented elsewhere.

Intrinsic alignments of galaxy ellipticities are another potential
source of systematic errors for cosmic shear measurement (see
\cite{BridleKing07} for the detail and references therein).  There are
two kinds of the contamination. The first is intrinsic-intrinsic galaxy
alignment (II) that may arise from neighboring galaxies residing in a
similar tidal field of large-scale structure
\cite{Mandelbaumetal06}. The second effect is a cross-correlation
between intrinsic ellipticity of a foreground galaxy and lensing
distortion of background galaxy shape (GI) because the foreground tidal
field affecting the intrinsic ellipticity of a foreground galaxy may
also cause lensing shear of a distant, background galaxy
(\cite{HirataSeljak04}). In general member galaxies of a cluster tend to
be more elliptical and therefore the width of the ellipticity histogram
(intrinsic ellipticity dispersion) will be smaller for cluster members
due to the absence of e.g. edge-on spirals.  Therefore if there are by
chance more clusters in a surveyed area then the noise on the shear
power spectrum will be slightly smaller. This is likely a tiny effect
and can be safely neglected.  Another interesting possibility is that if
there are many clusters in a surveyed area then both the II and GI
contamination to the cosmic shear power spectra would be larger, since
member galaxies of a cluster are more aligned with each other. Also the
stronger tidal field due to the cluster may also
cause the cluster members to be more 
anti-aligned 
with background galaxy
shapes due to lensing distortion by the cluster.  However, identifying
clusters within a given survey region could be a useful way to
remove/correct this II and GI contamination, which is another
interesting possibility of the combined cluster counts and cosmic shear
to use for future surveys.

For cluster counts, a most problematic source of the systematic errors
is the uncertainty in relating cluster observables to halo
mass.  One traditional way to tackle this obstacle is to investigate
properties of known massive clusters in great detail combing various
techniques (radio, optical, cluster lensing, X-ray and the SZ effect).
Or one 
might 
develop a reliable model for the mass-observable relation
using hydrodynamical simulations of cluster formation fully taking into
account the associated physical processes in the intracluster medium. Then
the mass-observable relation obtained in these ways could be used for
cluster counting statistics if the derived relation is a fair
representation of the mass-observable distribution of clusters in the
sample.

For lensing-based cluster counts, projection effects on a cluster
lensing signal due to mass along line-of-sight that is not associated
with the cluster introduce additional statistical errors in the mass
estimates of individual clusters \cite{WhiteVanWaerbeke02,HTY}.  In
addition, the scatter will be correlated with the cosmic shear power
spectrum, which we have
also 
 not taken into account.  To estimate the mass
estimate uncertainty and the effect of the ignored correlation in a
quantitative way, ray-tracing simulations of cosmic shear including
cluster lensing contributions will be needed. Also in practice
traditional methods (optical, X-ray, the SZ effect) will need to be
combined to exclude false clusters from the sample. These issues are
beyond the scope of this paper.

One may develop a
model to describe the 
mass-observable relation 
in terms of nuisance
parameters. Then  we could use cluster
observables available from a given survey to `self-calibrate',
i.e. determine both the cosmological parameters and the nuisance
parameters concurrently. In particular, it was shown in
\cite{MajumdarMohr04,LimaHu04} that adding the two-point correlation
function of clusters to cluster counts, both of which are drawn from the
same survey region, can be a 
useful way to self-calibrate the model
systematic errors in the mass-observable relation because the amplitude
of the cluster two-point function is very sensitive to halo bias that is
fairly well specified by halo masses.

Having the discussion above in mind, it would be interesting to address
whether the self-calibration regime could be attained for the combined
measurements of cluster counts and cosmic shear tomography, taking into
account the effects of systematic errors involved in each observable.
The cluster counts and cosmic shear depend on the cosmological
parameters in different ways and are sensitive to different systematic
errors. Hence one can use the combined measurement to constrain
simultaneously the cosmological parameters as well as the nuisance
parameters of systematic errors, mitigating degradation in the
cosmological parameter determination due to the systematic errors. Also
importantly
one could
realize, for a given
survey, the requirements on the control of the systematic errors
(photo-$z$, mass-observable relation etc) to attain the desired accuracy
of constraining dark energy parameters.  In this
direction, the cross-covariances between the cluster counts and cosmic
shear tomography may play an intriguing role, because (1) the
cross-correlations are cosmological signals arising from the cosmic mass
density field in large-scale structures or, in other words, there is in
general little 
cross-correlation between the systematic errors in the two
observables, and (2) a CDM structure formation model provides accurate
predictions for the cosmological cross-covariances. Hence including the
cross-covariances in the parameter estimations
 may be used as another viable monitor of the systematic
errors.  This interesting issue is beyond the scope of this paper and
will be presented elsewhere.

\section{Conclusion and Discussion}
\label{discuss}

In this paper we have estimated accuracies on cosmological parameters
derivable from a joint experiment of cluster counts and cosmic shear power
spectrum tomography when the two are drawn from the same survey
region. In doing this we have properly taken into account the
cross-covariance between the two observables, which describes how the
two observables are correlated in redshift and multipole space.  This is
necessary because the two experiments probe the same cosmic density
fields.
However note that, since we have ignored possible systematic errors, all
the results shown in this paper demonstrate pure cosmological powers for
the combined method.
We will below summarize our findings, and then will discuss the
remaining issues.

We have developed a formulation to compute the cross-covariance between the
cluster counts and the cosmic shear power spectra based on the dark matter
halo approach within the framework of a CDM structure formation model
(see Appendix).  The cross-covariance arises from the three-point
correlation function between the cluster distribution and two points of
the mass density fields.
It is found that there is a significant {\em
positive} cross-correlation between the cluster counts probing clusters
with masses $M\simgt 10^{14}M_\odot$ and the lensing power spectrum
amplitudes at multipoles $l\simgt 10^3$. Here the term `positive' is
used to mean that if fewer or more massive clusters are
found from a given survey region than the ensemble average, the lensing
power spectra will most likely have smaller or larger amplitudes, respectively.
The cross-correlation on angular and mass scales of interest arises
mainly from the 1-halo term contribution of the three-point
correlations: the correlation between one point within a given cluster
and the shearing effects on two different
 background galaxies due to the same cluster.
Our results are more accurate than the earlier work presented in
\cite{FangHaiman06}, because their work ignored the 1-halo term
contribution to the cross-covariance and only included the 2-halo term
contribution, which is dominant only on large angular scales where the
useful cosmological information can not be extracted.

To quantify the impact of the cross-covariance, we first investigated
the total signal-to-noise ($S/N$) ratios for a joint measurement of the
cluster counts and the lensing power spectrum.
It was
shown that an inclusion of the cross-covariance leads to degradation
and, depending on the mass thresholds or the lensing detection thresholds,
improvement in the $S/N$ ratios up to $\sim \pm
20\%$ compared to the case that the two observable are considered to be
independent (see Figs.~\ref{fig:sn} and \ref{fig:sn-wlsn}).
The improvement occurs when the cluster counts including
massive halos $M\simgt 10^{15}M_\odot$ are combined with the lensing
power spectrum measurement (also see
\cite{Neyrincketal06,NeyrinckSzapudi07} for the related discussion).
This occurs even though the $S/N$ ratio for the cluster
counts alone is much less than that for the lensing power spectrum
alone.
That is,
a knowledge of the number of such massive clusters for a given survey region
helps improve accuracies of the joint measurement. This improvement is
achievable only if the cross-covariance is {\em a priori} known by using
the theoretical predictions or by directly estimating the
cross-correlation from the survey region. We also note
that the results change greatly if we ignore the non-Gaussian error
contribution to the lensing power spectrum covariance, which arises from
the lensing trispectrum (see Fig.~\ref{fig:sn_1halo}).  This implies
that the lensing fields are highly non-Gaussian (see \cite{TJ07} for an
extensive discussion).

We then presented forecasts for accuracies of
the cosmological parameter determination for
the joint experiment. To do this we included redshift
binning for both the cluster counts and the lensing power spectrum,
motivated by the fact that the additional redshift information is very
useful to tighten the cosmological parameter constraints, especially the
dark energy parameters.  In this paper we considered two simplified
cluster selection criteria: one is a mass-selected cluster sample, and the other is
the lensing-based cluster sample, where the latter contains clusters
having the lensing signal greater than a given threshold in the sample.
For the mass-selected cluster counts, it was found that combining the
cluster counts and the lensing tomography leads to significant
improvement in the errors on the dark energy parameters by $\sim 40\%$
only if the cluster counts including
down to
less massive halos such as $M\simgt
10^{14}M_\odot$ are considered (see Fig.~\ref{fig:errors}).
The improvement is due to different
dependence of the two observables on the cosmological parameters.

On the other hand, for the lensing-based cluster counts, adding the
cluster counts to the lensing power spectrum tomography is more
complementary to tighten the errors on the dark energy parameters than the
mass-selected cluster counts (see Fig.~\ref{fig:errors-wlsn}).
For example, adding the counts of clusters with
the high lensing signals
$(S/N)_{\rm cluster}\simgt 6$  improves the dark
energy errors by a factor of 2, even though the counts contain many fewer
clusters
and probe a narrower redshift range
than the mass-selected clusters
of $M\simgt 5\times 10^{14}M_\odot$
 (see Fig.~\ref{fig:dndz-wl}).
 This result is
encouraging because such massive halos are rare and therefore it seems
relatively easy to make follow-up observations, e.g., in order to obtain
well-calibrated relations between cluster mass and observables (also see
\S~\ref{sys} for the discussion).  The reason lensing-based cluster counts
are more powerful is ascribed to the fact that the cluster lensing
signal itself depends on the cosmological parameters via the lensing
efficiency and the dependence amplifies the sensitivity of the cluster
counts to the dark energy parameters.
However, with low detection thresholds such as $(S/N)_{\rm
cluster}\simlt 3$ the lensing-based counts begin to suffer too
much from projection effects due to large-scale structures that are not
associated with the cluster.  Hence, if traditional mass-selected
cluster counts can go to lower masses then they might catch up with, or
overtake, the lensing-based counts in their constraining power.

For the impact of the cross-covariance on the parameter determination,
the effect is generally small for both the mass-selected and
lensing-based cluster counts.
This is partly because the lensing power spectra are
sensitive to the total number of clusters
\emph{roughly weighted by the cluster mass squared} whereas for
the cluster counts we simply added up the number of clusters
(see Appendix~\ref{appendix:toy-wl}).
This means that the two probes are not measuring such a similar
quantity and the cross-covariance is smaller than if they both
measured the unweighted total number of clusters.
Further, the redshift weighting is different for the lensing
power spectra and the cluster counts, so not all the halos are
in common.
It is also partly
a result of working in
multi-dimensional parameter space (9 parameters for our case).
Yet,
it is intriguing to note that the dark energy parameters are in most cases
improved by including the cross-covariance (see the lower panels of each
plot in Figs.~\ref{fig:errors} and \ref{fig:errors-wlsn}; also see
\cite{TJ07} for the related discussion).
 In summary, a joint experiment
of cluster counts and lensing power spectrum tomography will be
worth exploring in order to exploit full information on the cosmological
parameters from future massive surveys, and including the
cross-covariance will be needed in order
to correctly estimate the error bars.

In this work we have assumed that cluster counts measure the total
number of clusters above some threshold, in a number of redshift
bins.
In principle subdividing
cluster counts in mass or lensing signal bins
could
also  improve cosmological parameter constraints
\citep{MarianBernstein06,Hu03,Rozoetal07}. This would make the improvement
on including cluster counts to lensing power spectra even more
impressive, however the covariance may be more important than
we find in this paper.


Finally, we comment on a possibility for ultimate experiments combining
all observables available from one survey region. As we have
shown, one can combine different observables to improve accuracies of
the cosmological parameter determination, even though the observables
probe the {\em same} cosmic density fields. Besides the cluster counts
and the lensing power spectrum considered in this paper, there will be
other various observables available: cosmic shear bispectra or more
generally $n$-point correlation functions of the cosmic shear fields
\cite{TJ04}, $n$-point correlation functions of cluster and galaxy
distributions \cite{HuJain04}, small-scale cluster lensing signals
\cite{JainTaylor03},
cosmic flexion correlation functions \cite{BaconGoldbergRoweTaylor06,Okuraetal06}
and so on.  Then, one natural question raises: Can
we combine all the observables in order to improve the parameter
constraints as much as possible?  Or, in the presence of the
systematic errors, is there an optimal combination of the observables to
maximize the parameter constraints as well as most mitigate degradation
in the parameter constraints
due the systematic errors.
However, to address this interesting issue
quantitatively, all the covariances between the observables used have to
be correctly taken into account.
We believe that the formulation developed in
this paper would be useful to compute the
covariances for any observables and the combinations.  This kinds of
study will be worthwhile exploring in order to exploit the full
potential of future expensive surveys for constraining the nature of
mysterious dark energy components
and possible modifications of gravity.

\bigskip

{\it Acknowledgments} We thank G.~Bernstein, T.~Hamana, B.~Jain, M.~Jarvis,
R.~Sheth and J.~Weller for helpful discussions.
We also thank
anonymous referees for useful comments which led to improvement on the
manuscript.
M.~T.~ acknowledges the warm
hospitality of Astrophysics Group of University College London, where
this work was initiated.  M.~T.~ was in part supported by Grant-in-Aid
for the 21st Century COE ``Exploring New Science by Bridging
Particle-Matter Hierarchy'' at Tohoku University as well as
Grants-in-Aid for Scientific Research (Nos. 17740129 and 18072001) from
the Ministry of Education, Culture, Sports, Science and Technology of
Japan.  S.~B.~acknowledges the support of a Royal Society University
Research Fellowship.

\onecolumngrid
\appendix
\begin{center}
  {\bf APPENDIX}
\end{center}

In this Appendix, we describe a detailed formulation to
compute covariances between measurements of cluster counts, of
lensing power spectra and for the joint experiment. These are needed to
quantify the measurement errors and the error correlations between
different redshift and multipole bins for given survey parameters and
cosmological models.  The covariances are specifically predictable using
a secure model of non-linear gravitational clustering in structure
formation. To do this, we use the halo model approach developed in
\cite{TJ03a,TJ03b} (also see 
\cite{Peebles74,McClellandSilk77,ScherrerBertschinger91,Scoccimarroetal01}; 
\cite{CooraySheth} for a thorough review of the halo model).

\section{Halo model approach}

\subsection{A modeling of mass and cluster distributions}
\label{app:cluster}

In the halo model approach, we assume that all the matter is in halos
with density profile $\rho_h(\bmf{x}; m)$ that is parametrized by a mass
$m $ (e.g., the virial mass). In this setting, the mass density field at
an arbitrary spatial position $\bm{x}$, $\rho(\bmf{x})$, is written as
\begin{equation}
\rho_{\rm m}(\bmf{x})=\sum_i~ \rho_h(\bmf{x}-\bmf{x}_i; m_i)=
\sum_i~ m_i u_{m_i}(\bmf{x}-\bmf{x}_i)=\sum_i
\int\!dm~ \int\!d^3\bmf{x}' ~ \delta_D(m-m_i)\delta_D^3(\bmf{x}'-\bmf{x}_i)
m u_m(\bmf{x}-\bmf{x}'),
\label{eqn:dens}
\end{equation}
where we have introduced the normalized halo density profile
$u_{m}(\bmf{x})$ through $\rho_h(\bmf{x}; m)=m u_m(\bmf{x})$, the
summation $\sum_i$ runs over halos (the index $i$ denotes the $i$-th
halo) and
$\delta_D(\bm{x})$
is the Dirac delta function.  The first
equality just says that the mass density field at the position
$\bmf{x}$ is expressed as a superposition of density profiles of all
halos existing in the universe, where the vector between the position
$\bmf{x}$ and the halo position is given by $\bmf{x}-\bmf{x}_i$ with
$\bmf{x}_i$ being the $i$-th halo's center.

The ensemble average
over realisations of the universe
of the mass density field is shown to be
\begin{equation}
\skaco{\rho(\bmf{x})}=\int\!dm~ mn(m)\int\!d^3\bmf{x}'~ u_m(\bmf{x}-\bmf{x}')
=\int\!dm~ mn(m)=\bar{\rho}_m
\label{eqn:rhom}
\end{equation}
where we have assumed the ensemble average
$\skaco{\sum_i\delta_D(m-m_i)\delta^3_D(\bm{x}-\bm{x}_i)}=n(m)$ so that
the ensemble average does not depend on any specific spatial position.
Eq.~(\ref{eqn:rhom}) thus demonstrates that the ensemble average of the
mass density field is equal to the cosmic average mass density
$\bar{\rho}_m$ as expected.  Note that the mass function
$n(m)$  given in Press
\& Schechter \cite{PressSchechter} or the improved one in
\cite{ShethTormen} by definition satisfy the normalization condition
$\int\!dm~ n(m)(m/\bar{\rho}_{\rm m})=1$.
To properly define the halo mass $m$ for an extended halo profile the
the normalization condition $\int\!d^3\bm{x}~ u_m(\bm{x})=1$
must be
satisfied.
In this paper we assume the density is zero outside the virial radius
(see \cite{TJ03a} for
a detailed discussion about which mass definitions are
self-consistent in the halo model approach).

In this paper we consider constraints on cosmology calculated from
number counts of clusters in a hypothetical cluster
experiment (see \S~\ref{cl}).  In number counts the cluster
distribution is treated as points, and in other words one does not care
about the shape of the mass distribution within a cluster.  The relevant
quantity is the number density field of clusters, which can be
straightforwardly modeled based on the halo model formulation, from a
slight modification of Eq.~(\ref{eqn:dens}):
\begin{equation}
n_{\rm cl}(\bmf{x})=\sum_i~ \delta_D^3(\bmf{x}-\bmf{x}_i)S(m_i)
=\sum_i\int\!\!dm~ S(m)\delta_D(m-m_i)\int\!\!d^3\bmf{x}' \delta_D^3(\bmf{x}'-\bmf{x}_i)
\delta_D^3(\bmf{x}-\bmf{x}')
\label{eqn:ndens}
\end{equation}
where the subscript `cl' in $n_{\rm cl}$ stands for cluster and
$S(m)$
denotes the selection function that discriminates clusters used in the
number counts from other halos.  The ensemble average of the cluster
number density field (\ref{eqn:ndens}) is found, similarly to in
Eq.~(\ref{eqn:rhom}), to be given by
\begin{equation}
\bar{n}_{\rm cl}=\skaco{n_{\rm cl}(\bmf{x})}=
\int\!\!dm~ n(m)S(m)\int\!\!d^3\bmf{x}'~
\delta_D^3(\bmf{x}-\bmf{x}')=\int\!\!dm~ n(m)S(m).
\label{eqn:avendens}
\end{equation}
Thus, as expected, the ensemble average of the number density field is
indeed given by mass-integral of the halo mass function with a given
selection criterion.

\subsection{Correlation functions of mass and cluster distributions}
\label{clcounts}

In this subsection we use the halo model approach to
derive the correlation
functions of the cluster distribution and the cross-correlation between the
cluster distribution and the mass distribution.
These are needed to quantify
covariances of the cluster counts and the cross-covariance between the
cluster counts and the lensing power spectrum, respectively.

From Eq.~(\ref{eqn:ndens}), the 2-point correlation function of the
cluster number density field can be computed as
\begin{eqnarray}
\bar{n}_{\rm cl}^2\xi_{cc}(|\bm{x}_1-\bm{x}_2|)&\equiv&
\skaco{n_{\rm cl}(\bm{x}_1)n_{\rm
cl}(\bm{x}_2)}~ -\bar{n}_{\rm cl}^2
\nonumber\\
&=&\kaco{\sum_iS^2(m_i)\delta^3_D(\bm{x}_1-\bm{x}_i)\delta_D^3(\bm{x}_2-\bm{x}_i)
}+\kaco{\sum_{\stackrel{i,j}{{i\ne j}}}
S(m_i)S(m_j)\delta^3(\bm{x}_1-\bm{x}_i)
\delta^3_D(\bm{x}_2-\bm{x}_j)}
\nonumber\\
&=&
\kaco{\sum_{i}\int\!dm\int\!d^3\bm{y}~
S(m)\delta_D(m-m_i)\delta^3_D(\bm{x}_1-\bm{y})\delta^3_D(\bm{x}_2-\bm{y})
\delta^3_D(\bm{y}-\bm{x}_i)
}
\nonumber\\
&&+
\left\langle\sum_{\stackrel{i,j}{ i\ne j}}\int\!dm
\int\!d^3\bm{y}~
S(m)
\delta_D(m-m_i)
\delta^3_D(\bm{x}_1-\bm{y})\delta^3_D(\bm{y}-\bm{x}_i)\right.\nonumber\\
&&\hspace{3em}\times \left. \int\!dm' \int\!d^3\bm{y}'~
S(m')\delta_D(m'-m_j)
\delta^3_D(\bm{x}_2-\bm{y}')
\delta^3_D(\bm{y}'-\bm{x}_j)
\right\rangle
\nonumber\\
&=&
\int\!\!dm~ n(m)S(m) \int\!\!d^3\bmf{y}~
 \delta_D^3(\bmf{x}_1-\bmf{y})
\delta_D^3(\bmf{x}_2-\bmf{y})
\nonumber\\
&&+\int\!\!dm n(m)S(m) \int\!\!d^3\bmf{y}\delta_D^3(\bmf{x}_1-\bmf{y})
\int\!\!dm' n(m')S(m') \int\!\!d^3\bmf{y}'\delta_D^3(\bmf{x}_2-\bmf{y}')
\xi_h(\bmf{y}-\bmf{y}'; m,m')\nonumber\\
&=&\bar{n}_{\rm cl}\delta_D^3(\bmf{x}_1-\bmf{x}_2)
+\left[
\int\!\!dm~ n(m)S(m)b(m)\right]^2
\xi^L_\delta(\bmf{x}_2-\bmf{x}_1),
\label{eqn:cl2pt}
\end{eqnarray}
where we have used $S^2(m_i)=S(m_i)$ since $S(m_i)=1$ or $0$ (see text
below Eq.~[\ref{eqn:ncl}]), and the 2-point correlation function
$\xi_{cc}$ is dimension-less as usual. The first term on the
r.h.s. represents the 1-halo term contribution that arises due to
discrete nature of clusters probed (see \S~31 in \cite{Peebles80}),
while the second term gives the 2-halo term contribution that arises
from clustering
of
clusters.  Note that in
the last line on the r.h.s. we have assumed the 2-point correlation
between different two halos with masses $m$ and $m'$ is given by the
linear theory mass correlation function, $\xi_\delta^L(r)$, multiplied by the
halo bias parameters $b(m)$ and $b(m')$: $\xi_h(r;
m,m')=b(m)b(m')\xi^L_\delta(r)$.

Next we consider the 3-point correlation functions between the cluster
number density field and
{\em two} points on the mass density fluctuation field.
This 3-point function is needed to quantify the cross-covariance
between the cluster counts and the lensing power spectrum, where the
lensing power spectrum arises from the 2-point correlation of the mass
fluctuation field.  The 3-point correlation function we are interested
in is defined as
\begin{eqnarray}
\bar{n}_{\rm cl}\zeta_{c\delta\delta}(\bm{x}_1,\bm{x}_2,\bm{x}_3)&\equiv&
\skaco{\delta\!n_{\rm cl}(\bmf{x}_1)\delta_m(\bmf{x}_2)\delta_m(\bmf{x}_3)}
\nonumber\\
&=&\bar{n}_{\rm cl}\left[\zeta^{1h}_{c\delta\delta}(x_1,x_2,x_3)
+\zeta^{2h}_{c\delta\delta}(x_1,x_2,x_3)
+\zeta^{3h}_{c\delta\delta}(x_1,x_2,x_3)
\right],
\label{eqn:zetacdd}
\end{eqnarray}
where $\zeta_{c\delta\delta}$ is dimension-less, $\delta_m$ denotes the
mass density fluctuation field, and $\delta\!n_{\rm cl}(\bm{x}_1)$ is
the fluctuation part of the cluster number density field $n_{\rm
cl}(\bm{x})$ in Eq.~(\ref{eqn:ndens}) (the homogeneous part does not
contribute to the 3-point correlation).  In the second equality on the
r.h.s. we divided the 3-point function into three distinct
contributions: the 1-, 2- and 3-halo terms
from the picture of the halo model approach.  Note that the 3-point
correlation function is given as a function of triangle configuration,
and the amplitude is invariant under parallel translation and rotational
transformation of triangle configuration assuming the statistical
symmetry. Therefore, the 3-point correlation function is specified by
three parameters that describe a triangle configuration, e.g. three side
lengths.

Using a similar calculation procedure as used in
Eq.~(\ref{eqn:cl2pt}), the 1-halo term contribution to
$\zeta_{c\delta\delta}$ can be computed as
\begin{eqnarray}
\bar{n}_{\rm
 cl}\zeta_{c\delta\delta}^{1h}
 (\bm{x}_1,\bm{x}_2,\bm{x}_3)
&=&\int\!\!dm~ n(m) S(m)
\int\!\!d^3\bmf{x}_1'~ \delta_D^3(\bmf{x}_1-\bmf{x}_1')
\frac{m}{\bar{\rho}_m}
u_m(\bmf{x}_2-\bmf{x}_1')\frac{m}{\bar{\rho}_m}
u_m(\bmf{x}_3-\bmf{x}_1')
\nonumber\\
&=& \int\!\!dm~ n(m)\left(\frac{m}{\bar{\rho}_m}\right)^2
S(m)
u_m(\bmf{x}_2-\bmf{x}_1)u_m(\bmf{x}_3-\bmf{x}_1).
\label{eqn:zeta1h}
\end{eqnarray}

The 2-halo term contribution $\zeta_{c\delta\delta}^{2h}$ is found to be
\begin{eqnarray}
\bar{n}_{\rm cl}\zeta^{2h}_{c\delta\delta}(\bm{x}_1,\bm{x}_2,\bm{x}_3)
&=&\int\!\!dm_1
 n(m_1)S(m_1)b(m_1)\int\!\!dm_2
n(m_2)b(m_2)\! \left(\frac{m_2}{\bar{\rho}_m}\right)^2
\!\int\!\!d^3\bmf{y}
u_{m_2}(\bmf{x}_2-\bmf{y})
u_{m_2}(\bmf{x}_3-\bmf{y})
\xi^L_\delta(\bmf{x}_1-\bmf{y})\nonumber\\
&&\hspace{-4.5em}+\int\!\!dm_1~
 n(m_1)b(m_1)S(m_1)
\frac{m_1}{\bar{\rho}_m}
u_{m_1}(\bmf{x}_2-\bmf{x}_1)
\int\!\!dm_2~
n(m_2)b(m_2)
\frac{m_2}{\bar{\rho}_m}
\int\!\!d^3\bmf{y}~ u_{m_2}(\bmf{x}_3-\bmf{y})
\xi^L_\delta(\bmf{x}_1-\bmf{y})\nonumber\\
&&\hspace{-4.5em}+\int\!\!dm_1~
 n(m_1)b(m_1)S(m_1)
\frac{m_1}{\bar{\rho}_m}
u_{m_1}(\bmf{x}_3-\bmf{x}_1)
\int\!\!dm_2~
n(m_2)b(m_2)
\frac{m_2}{\bar{\rho}_m}
\int\!\!d^3\bmf{y}~
u_{m_2}(\bmf{x}_2-\bmf{y})\xi^L_\delta(\bmf{x}_1-\bmf{y})
\label{eqn:zeta2h}
.
\end{eqnarray}
The 2-halo term arises from the correlation between two different halos,
where the clustering strength between the two halos is given by
$b(m_1)b(m_2)\xi^L_\delta$ the same as we did in Eq.~(\ref{eqn:cl2pt}).

The 3-halo term $\zeta^{3h}_{c\delta\delta}$ in Eq.~(\ref{eqn:zetacdd}),
which arises from the correlation between three different halos,
is given by
\begin{eqnarray}
\bar{n}_{\rm cl}\zeta^{3h}_{c\delta\delta}(\bm{x}_1,\bm{x}_2,\bm{x}_3)
&=&\int\!\!dm_1~
 n(m_1)S(m_1)b(m_1)
\int\!\!dm_2~ n(m_2)b(m_2)\frac{m_2}{\bar{\rho}_m}\int\!\!d^3\bmf{y}~
u_{m_2}(\bmf{x}_2-\bmf{y})\nonumber\\
&&\times
\int\!\!dm_3~ n(m_3)b(m_3)\frac{m_3}{\bar{\rho}_m}
\int\!\!d^3\bmf{y}'~
u_{m_3}(\bmf{x}_3-\bmf{y}')\zeta^{PT}_\delta(\bmf{x}_1,\bmf{y},\bmf{y}'),
\label{eqn:zeta3h}
\end{eqnarray}
where $\zeta^{\rm PT}_\delta$ is the perturbation theory prediction for
the 3-point correlation function of the mass density field, and we have
assumed that the 3-point correlation of halo distribution can be
expressed by $\zeta^{\rm PT}_\delta$ multiplied by halo bias
parameters: $\zeta_h(\bm{x}_1,\bm{x}_2,\bm{x}_3; m_1,m_2,m_3)=b(m_1)
b(m_2)b(m_3)\zeta^{\rm PT}_\delta(\bm{x}_1,\bm{x}_2,\bm{x}_3)
$.

For convenience for the following discussion, we derive the
Fourier-transformed counterpart of the 3-point correlation function
$\zeta_{c\delta\delta}$, the bispectrum $B_{c\delta\delta}$. The
bispectrum is related to the 3-point correlation function via the
Fourier transform given by
\begin{equation}
\zeta_{c\delta\delta}(\bmf{x}_1,\bmf{x}_2,\bmf{x}_3)\equiv
\int\!\!\frac{d^3\bmf{k}_1}{(2\pi)^3}\frac{d^3\bmf{k}_2}{(2\pi)^3}
\frac{d^3\bmf{k}_3}{(2\pi)^3}B_{c\delta\delta}(\bmf{k}_1,\bmf{k}_2,\bmf{k}_3)
e^{i\bmf{k}_1\cdot\bmf{x}_1+i\bmf{k}_2\cdot\bmf{x}_2+i\bmf{k}_3\cdot\bmf{x}_3
}
(2\pi)^3\delta^3_D(\bmf{k}_1+\bmf{k}_2+\bmf{k}_3).
\label{eqn:bispdef}
\end{equation}
Note that the bispectrum is also defined by the ensemble average of the
three Fourier-transformed coefficients of the cluster number density
field and the mass density fields as
\begin{equation}
\skaco{\tilde{\delta n}(\bmf{k}_1)\tilde{\delta}_m(\bmf{k}_2)\tilde{\delta}_m(\bmf{k}_3)}
=
B_{c\delta\delta}(\bmf{k}_1,\bmf{k}_2,\bmf{k}_3)
(2\pi)^3\delta^3_D(\bmf{k}_1+\bmf{k}_2+\bmf{k}_3).
\end{equation}
Similarly to in Eq.~(\ref{eqn:zetacdd}), the bispectrum can be divided
into the 1-, 2- and 3-halo term contributions as
\begin{equation}
B_{c\delta\delta}=B^{1h}_{c\delta\delta}
+B^{2h}_{c\delta\delta}+B^{3h}_{c\delta\delta}.
\label{eqn:bisphalos}
\end{equation}

Combining  Eq.~(\ref{eqn:zeta1h}) and Eq.~(\ref{eqn:bispdef}),
 the 1-halo term of the bispectrum, $B^{1h}_{c\delta\delta}$, is found to
be
\begin{equation}
\bar{n}_{\rm cl}B_{c\delta\delta}^{1h}(\bmf{k}_1,\bmf{k}_2,\bmf{k}_3)=
\int\!\!dm~ n(m)\left(\frac{m}{\bar{\rho}_m}\right)^2
S(m) \tilde{u}_m(k_2)\tilde{u}_m(k_3).
\label{eqn:bisp1h}
\end{equation}
Here $\tilde{u}_m(k)$ is the Fourier transform of the halo density
profile defined as
\begin{equation}
\tilde{u}_m(k)\equiv \int^{r_{\rm vir}}_0 4\pi r^2dr~ u(r)j_0(kr)
\label{eqn:um}
\end{equation}
where we have assumed a spherically symmetric density profile for
simplicity, $r_{\rm vir}$ is the virial radius (more generally, the
boundary radius of a halo used for the mass definition), and $j_0(x)$ is
the zero-th order spherical Bessel function, $j_0(x)=\sin x/x$.  Note
that $\tilde{u}_m$ has the property that $\tilde{u}_m(k)=1$ for
$k\rightarrow0$.  From Eq.~(\ref{eqn:bisp1h}), one finds that the
bispectrum does not depend on wavenumber $\bm{k}_1$ that comes from the
Fourier transform of the cluster number density contribution; since the
cluster distribution is modeled as discrete points, the contribution to
the 1-halo term arises from representative {\em one} point within a
given halo (e.g., the halo center), which corresponds to white noise
(therefore no $k$-dependence) in the Fourier transform.

Similarly, from Eq.~(\ref{eqn:zeta2h}), the 2-halo term contribution to
the bispectrum, $B_{c\delta\delta}^{2h}$, is given by
\begin{eqnarray}
\bar{n}_{\rm cl}B_{c\delta\delta}^{2h}(\bmf{k}_1,\bmf{k}_2,\bmf{k}_3)&=&
\left[
\int\!\!dm_1~ n(m_1)S(m_1)b(m_1)
\right]
\left[
\int\!\!dm_2~ n(m_2)b(m_2)\left(\frac{m_2}{\bar{\rho}_{\rm m}}\right)^2
\tilde{u}_{m_2}(k_2)\tilde{u}_{m_2}(k_3)\right]
P^L_\delta(k_1)\nonumber\\
&&+\left[
\int\!\!dm_1~ n(m_1)S(m_1)b(m_1)\frac{m_1}{\bar{\rho}_m}
\tilde{u}_{m_1}(k_2)
\right]
\left[
\int\!\!dm_2~ n(m_2)b(m_2)\frac{m_2}{\bar{\rho}_{\rm m}}
\tilde{u}_{m_2}(k_3)
\right]
P^L_\delta(k_3)\nonumber\\
&&+\left[
\int\!\!dm_1~ n(m_1)S(m_1)b(m_1)\frac{m_1}{\bar{\rho}_m}
\tilde{u}_{m_1}(k_3)
\right]
\left[
\int\!\!dm_2~ n(m_2)b(m_2)\frac{m_2}{\bar{\rho}_{\rm m}}
\tilde{u}_{m_2}(k_2)
\right]
P^L_\delta(k_2).
\label{eqn:bisp2h}
\end{eqnarray}
Here, square brackets are used in order to emphasize that the halo mass
integral in the terms enclosed by square brackets can be calculated
separately from other terms.

From Eq.~(\ref{eqn:zeta3h}), the 3-halo term of the bispectrum,
$B_{c\delta\delta}^{3h}$, is found to be
\begin{eqnarray}
\bar{n}_{\rm cl}B_{c\delta\delta}^{3h}(\bmf{k}_1,\bmf{k}_2,\bmf{k}_3)&=&
\left[
\int\!\!dm_1~ n(m_1)S(m_1)b(m_1)
\right]
\left[
\int\!\!dm_2~ n(m_2)b(m_2)\frac{m_2}{\bar{\rho}_{\rm m}}
\tilde{u}_{m_2}(k_2)\right]
\nonumber\\
&&\hspace{5em}\times\left[
\int\!\!dm_3~ n(m_3)b(m_3)\frac{m_3}{\bar{\rho}_{\rm m}}
\tilde{u}_{m_3}(k_3)\right]
B_\delta^{PT}(\bmf{k}_1,\bmf{k}_2,\bmf{k}_3),
\label{eqn:bisp3h}
\end{eqnarray}
where $B^{\rm PT}_\delta$ is the perturbation theory prediction for the mass
bispectrum given by
\begin{equation}
B_\delta^{\rm PT}(\bmf{k}_1,\bmf{k}_2,\bmf{k}_3)=\left[
\frac{10}{7}+\left(\frac{k_1}{k_2}+\frac{k_2}{k_1}\right)\left(\frac{\bmf{k}_1\cdot\bmf{k}_2}{k_1k_2}\right)
+\frac{4}{7}\frac{(\bmf{k}_1\cdot\bmf{k}_2)^2}{k_1^2k_2^2}
\right]P^L_\delta(k_1)P^L_\delta(k_2)+ \mbox{2 perm.},
\label{eqn:bisppt}
\end{equation}
where the terms denoted by `2 perm.' are obtained from two permutations
of $\bm{k}_1\leftrightarrow \bm{k}_3$ and $\bm{k}_2\leftrightarrow
\bm{k}_3$
(e.g., see
by \cite{Bernardeauetal02} for the derivation of $B_\delta^{\rm PT}$).

\section{Covariances of  the cluster number counts and lensing power spectrum }
\label{appendix:cov}

Using the correlation functions shown in the preceding section, we
are ready to derive covariances of the cluster counts, and the
cross-covariance with the lensing power spectrum.

\subsection{Covariances of the cluster counts}
\label{appendix:clcov}

In this paper we have considered the average angular number density of
clusters drawn from a given survey region on the sky as our observable
from the cluster count experiments. An estimator of the angular number density
is given by Eq.~(\ref{eqn:angcl}), which is slightly modified from
Eq.~(\ref{eqn:ndens}) so that the counts include redshift binning via a
modification of the selection function to $S_{(b)}(m; z)$ (the subscript
`$b$' denotes the $b$-th redshift bin). The covariance between the
number densities in redshift bins $b$ and $b'$ is defined by
Eq.~(\ref{eqn:clcov-def}) and can be computed, using
Eq.~(\ref{eqn:cl2pt}) and Limber's approximation, as
\begin{eqnarray}
[\bm{C}]^c_{bb'}&\equiv&
\skaco{{\cal N}_{(b)}{\cal
 N}_{(b')}}-N_{(b)}N_{(b')}
\nonumber\\
&=&
\int\!\!d^2\bmf{\theta}W(\bmf{\theta})
\int\!\!d^2\bmf{\theta'}W(\bmf{\theta}')
\int\!\!d\chi\frac{d^2V}{d\chi d\Omega}
\int\!\!d\chi'\frac{d^2V}{d\chi' d\Omega}
\left[
\skaco{n_{{\rm cl},(b)}(\chi,\chi\bmf{\theta})n_{{\rm
cl}, (b')}(\chi',\chi'\bmf{\theta}')}-\bar{n}_{{\rm cl}, (b)}\bar{n}_{{\rm cl}, (b')}
\right]
\nonumber\\
&=&\int\!\!d^2\bmf{\theta}W(\bmf{\theta})
\int\!\!d^2\bmf{\theta'}W(\bmf{\theta}')
\int\!\!d\chi\frac{d^2V}{d\chi d\Omega}
\int\!\!d\chi'\frac{d^2V}{d\chi' d\Omega}
\left[\delta_D(\chi-\chi')
\delta^2_D(\chi\bm{\theta}-\chi'\bm{\theta}')\int\!dm~ S_{(b)}(m;z)
S_{(b')}(m;z')n(m)
\right.\nonumber\\
&&
+\left.\left\{\int\!dm_1~ S_{(b)}(m_1;z)n(m_1)b(m_1)\right\}
\left\{\int\!dm_2~ S_{(b')}(m_2;z')n(m_2)b(m_2)\right\}
\xi^L_\delta(\bm{x}-\bm{x}';z,z')
\right]\nonumber\\
&\approx&\int\!\!d^2\bmf{\theta}W^2(\bmf{\theta})
\int\!\!d\chi
\left(\frac{d^2V}{d\chi d\Omega}\right)^2\chi^{-2}\int\!dm~ S_{(b)}(m;z)S_{(b')}(m,z)n(m)
\nonumber\\
&&+\delta^K_{bb'}\int\!\!d^2\bmf{\theta}W(\bmf{\theta})
\!\int\!\!d^2\bmf{\theta'}W(\bmf{\theta}')
\!\int\!\!d\chi\!
\left(\frac{d^2V}{d\chi d\Omega}\right)^2\!\!\chi^{-2}\!\!
\left[\int\!dm~ S_{(b)}(m;z)
n(m)b(m)\right]^2\!\!
\int\!\!\frac{d^2\bm{l}}{(2\pi)^2}P^L_\delta
\!\!\left(k=\frac{l}{\chi};\chi
\right)
e^{i\bm{l}\cdot(\bm{\theta}-\bm{\theta}')}\nonumber\\
&=&\delta^{K}_{bb'}\frac{N_{(b)}}{\Omega_{\rm s}}
+\delta^K_{bb'}\int\!\!d\chi
\left(\frac{d^2V}{d\chi d\Omega}\right)^2\chi^{-2}
\left[\int\!dm~ S_{(b)}(m;z)
n(m)b(m)\right]^2
\int\!\!\frac{ldl}{2\pi}P^L_\delta(k=l/\chi;
\chi)|\tilde{W}(l\Theta_s)|^2,
\label{eqn:clcov-calc}
\end{eqnarray}
where $P^L_\delta$ is the linear mass power spectrum, $N_{(b)}$ is the
ensemble average of the angular number density estimator given by
Eq.~(\ref{eqn:cc-Nb}), $\Omega_{\rm s}$ is the surveyed area, and
$\tilde{W}(l)$ is the Fourier transform of the survey window function
(see text below Eq.~[\ref{eqn:clcov}] for the details).  To be more
explicit for the calculation procedures above, in the fourth line on the
r.h.s., we have employed Limber's approximation for the calculation
of the 2-halo term: the multiple line-of-sight integral of clustering
contributions at different redshifts is replaced with the single
line-of-sight integral. This is a good approximation when the redshift
bin width of the number counts is sufficiently thicker than the
correlation length of clusters.
In addition, since
we assume the cluster redshift bins do not overlap
the selection
functions for two redshift bins have the property
$S_{(b)}S_{(b')}=\delta^{K}_{bb'}S_{(b)}^2=\delta_{bb'}^KS_{(b)}$.
Therefore the
1- and 2-halo terms are both proportional to the Kronecker delta
function $\delta^{K}_{bb}$, ensuring that there is no correlation
between the cluster counts in different redshift bins.  In the last
equality for the 1-halo term calculation, we have used $ (d^2V/d\chi
d\Omega)^2\chi^{-2}=d^2V/d\chi d\Omega $ and the integral of the survey
window function is computed as $\int\!\!d^2\bm{\theta}~
W^2(\bm{\theta})=1/\Omega_{\rm s}$ because of the normalization
condition for the window function,
$\int\!d^2\bm{\theta}W(\bm{\theta})=1$ (we have assumed a top-hat window
function for simplicity).

The covariance of the cluster counts has two distinct contributions; the
first term in Eq.~(\ref{eqn:clcov-calc}) represents the shot noise
contribution arising from the imperfect sampling of fluctuations due to
a finite number of clusters, while the second term represents the
sampling variance arising from fluctuations of the cluster distribution
due to a finite survey volume.  It should be stressed that, based on the
halo model approach, we can thus derive the shot noise contribution to
the covariance without {\it ad hoc} introducing the term as
conventionally done in the literature \cite{Haimanetal00}. We would like
to also emphasize that the sample variance in Eq.~(\ref{eqn:clcov-calc})
is consistent with the results derived in \cite{HuKravtsov03}.

\subsection{The lensing power spectrum covariance}
\label{appendix:wlcov}

In this subsection, we derive the covariance of the lensing power
spectrum following the formulation developed in \cite{TJ07}.
In this paper we have focused on the lensing power spectrum
as our lensing observable. Under a flat-sky approximation, the power
spectrum is constructed from the two-dimensional Fourier transform of
the measured convergence field available over a given survey
region.
The Fourier decomposition has to be done for modes
taken from a finite survey region. For such a finite sky measurement,
infinite number of the Fourier modes are not available. Therefore, the
Fourier decomposition is by nature discrete, and the fundamental mode is
limited by the size of surveyed area, $l_{\rm min} =
2\pi/\Theta_s$, where
the survey area is given by $\Omega_{\rm s}=\Theta_s^2$ (we assume a
square survey geometry for simplicity)
 \footnote{Exactly speaking we are not consistent
for a treatment of survey geometry, compared to our another
assumption of a circular
geometry for  the survey window function used in the cluster
counts (see around  Eq.~[\ref{eqn:clcov}]). However, most information of
the lensing power spectrum comes from small angular scales, so the
geometry does not affect our results as long as the survey area is sufficiently
large. In addition the lensing covariance depends on the survey area, not
on the survey geometry, to a zero-th order approximation.
For these reasons, we use the approximation for
computational simplicity.}.
For this case, the convergence
field can be expanded using the discrete Fourier decomposition as
\begin{equation}
\kappa(\bmf{\theta})=
\frac{1}{\Omega_{\rm s}}
\sum_{\bmf{l}}\tilde{\kappa}_{\bmf{l}}e^{i\bmf{l}\cdot\bmf{\theta}},
\label{eqn:four}
\end{equation}
where the summation runs over the combination of integers $(n_x,n_y)$ for
$\bmf{l}=(2\pi/\Theta_{\rm s})(n_x,n_y)$.  Here we consider the
convergence field for a single source redshift bin for simplicity, and
it is very straightforward to extend the following discussion to a
tomographic case.  For an infinite survey limit ($\Theta_{\rm
s}\rightarrow \infty$), the Fourier transform above becomes
\begin{equation}
\kappa(\bmf{\theta})=
\frac{1}{\Omega_{\rm s}}
\sum_{\bmf{l}}\tilde{\kappa}_{\bmf{l}}\rightarrow
\int\!\!\frac{d^2\bmf{l}}{(2\pi)^2}\tilde{\kappa}_{\bmf{l}}e^{i\bmf{l}\cdot\bmf{\theta}}.
\label{eqn:kappa-limit}
\end{equation}
In the discrete Fourier expansion, the orthogonal relation for eigenmode
function $e^{i\bmf{l}\cdot\bmf{\theta}}$ is given by
\begin{equation}
\int_{\Omega_{\rm
 s}}\!d^2\bmf{\theta}e^{i(\bmf{l}-\bmf{l}')\cdot\bmf{\theta}}
=\Omega_{\rm s}\delta^{K}_{\bmf{l}-\bmf{l}'}
\label{eqn:orth}
\end{equation}
where the integration range for $\int\!\!d^2\bm{\theta}$ is confined to
the survey area and $\delta^K_{\bmf{l}-\bmf{l}'}$ is the Kronecker type
delta function for vectors defined as
\begin{equation}
\delta^{K}_{\bmf{l}-\bmf{l}'}=
\left\{
\begin{array}{ll}
1 & \mbox{if $\bmf{l}=\bmf{l}'$} \\
0 & \mbox{otherwise}
\end{array}
\right..
\end{equation}
The orthogonal relation (\ref{eqn:orth}) implies that the Kronecker
delta function $\delta^K_{\bm{l}-\bm{l}'}$ should be replaced with the
Dirac delta function for an infinite survey $(\Theta_{\rm s}\rightarrow
\infty)$ (also see \cite{Hu00}) as
\begin{equation}
\Omega_{\rm s}\delta^{K}_{\bmf{l}-\bmf{l}'}\rightarrow
 (2\pi)^2\delta_D^2(\bmf{l}-\bmf{l}').
\end{equation}
From this relation, the definition for the convergence power spectrum is
also modified for the finite-sky Fourier decomposition to
\begin{equation}
\skaco{\tilde{\kappa}_{\bmf{l}_1}\tilde{\kappa}_{\bmf{l}_2}}=\Omega_{\rm
 s}
\delta^{K}_{\bmf{l}_1+\bmf{l}_2}P_\kappa(l_1)
\label{eqn:psdef}
\end{equation}
in that the power spectrum definition matches the conventional one
$\skaco{\tilde{\kappa}_{\bmf{l}_1}\tilde{\kappa}_{\bmf{l}_2}}
=(2\pi)^2\delta_D^2(\bmf{l}_1+\bmf{l}_2)P_\kappa(l_1)$ for an infinite
survey limit.  Note that, from Eqs.~(\ref{eqn:four}) and
(\ref{eqn:orth}), the inverse Fourier transform is given by
$
\tilde{\kappa}_{\bmf{l}}=\int_{\Omega_{\rm
 s}}\!\!d^2\bmf{\theta}~  \kappa(\bmf{\theta})e^{i\bmf{l}\cdot\bmf{\theta}}.
$

Once the discrete Fourier modes of the convergence field for a finite
survey are defined, an estimator for the convergence power spectrum
measurement may be defined as
\begin{equation}
P^{\rm est}_\kappa(l)=\frac{1}{\Omega_{\rm s}N_p(l)}\sum_{\bmf{l}; l\in l_b}
\tilde{\kappa}_{\bmf{l}}\tilde{\kappa}_{-\bmf{l}},
\label{eqn:psest_appendix}
\end{equation}
where the summation runs over all the Fourier modes whose length is in
the range of $l-\delta l/2\le |\bmf{l}| \le l+\delta l/2$ for a given
bin width $\delta l$. Here $N_p(l)$ is the number of modes taken for the
summation, and is given by $N_p(l)=\sum_{\bmf{l}_i; l\in l_b}\approx
2\pi l\delta l/(2\pi/\Theta_s)^2=2 l\delta lf_{\rm sky}$, where $f_{\rm
sky}$ is the sky coverage as $\Omega_{\rm s}=\Theta_s^2=4\pi f_{\rm
sky}$. Note that Eq.~(\ref{eqn:psest_maintext}) corresponds to an
integral form of Eq.~(\ref{eqn:psest_appendix}), where the two
approximately match each other for large $l\gg 1/\Theta_s$.

Using Eq.~(\ref{eqn:psdef}), the ensemble average of the power spectrum
estimator (\ref{eqn:psest_appendix}) is found to indeed give the
underlying true power spectrum $P_\kappa(l)$:
\begin{eqnarray}
\skaco{P_\kappa^{\rm est}(l)}&=&\frac{1}{\Omega_{\rm
 s}N_p(l)}\sum_{\bmf{l}; l\in l_b}\skaco{\tilde{\kappa}_{\bmf{l}}
\tilde{\kappa}_{-\bmf{l}}}= \frac{1}{N_p(l)} \sum_{\bmf{l}; l\in l_b}P_\kappa(l)\nonumber\\
&\approx& \frac{1}{N_p(l)}P_\kappa(l) \sum_{\bmf{l}; l\in l_b}
=P_\kappa(l)
\label{eqn:psest_vs_underlying}
\end{eqnarray}
where we have assumed that the power spectrum changes little within the
bin width in the third equality on the r.h.s.

The covariance between the convergence power spectra in multipole bins
$l$ and $l'$ is defined as
\begin{eqnarray}
[\bmf{C}^{g}]_{ll'}&\equiv&
{\rm Cov}[P^{\rm est}_{\kappa}
(l),P^{\rm est}_{\kappa}(l')]
=\skaco{P_\kappa^{\rm est}(l)P_\kappa^{\rm est}(l')}
-P_\kappa(l)P_\kappa(l')
\nonumber\\
&=& \frac{1}{\Omega_{\rm s}N_p(l)}\frac{1}{\Omega_{\rm s}N_p(l')}
\sum_{\bmf{l}; l\in l_b}\sum_{\bmf{l}'; l'\in l'_b}
\skaco{\tilde{\kappa}_{\bmf{l}}\tilde{\kappa}_{-\bmf{l}}\tilde{\kappa}_{\bmf{l}'}
\tilde{\kappa}_{-\bmf{l}'}}-P_\kappa(l)P_\kappa(l').
\label{eqn:pcovdef}
\end{eqnarray}
Thus an estimation of the power spectrum covariance requires a knowledge
on the 4-point correlation functions of the convergence field.  The
4-point correlation function generally has two contributions; one is the
Gaussian contribution given by the power spectrum, and the other is the
non-Gaussian contribution that is the connected part of the 4-point
function, the so-called trispectrum.  The trispectrum of the convergence
field is naturally induced by non-linear evolution of gravitational
clustering in structure formation, which carries additional information
beyond that of the 2-point functions.  For a finite-sky Fourier
decomposition, assuming the statistically isotropic, random field for $\tilde{\kappa}_{\bmf{l}}$,
the 4-point function in Eq.~(\ref{eqn:pcovdef}) can be
expressed in terms of the power spectrum and the trispectrum as
\begin{eqnarray}
\skaco{\tilde{\kappa}_{\bmf{l}}\tilde{\kappa}_{-\bmf{l}}\tilde{\kappa}_{\bmf{l}'}
\tilde{\kappa}_{-\bmf{l}'}}&=&
\skaco{\tilde{\kappa}_{\bmf{l}}\tilde{\kappa}_{-\bmf{l}}}
\skaco{\tilde{\kappa}_{\bmf{l}'}\tilde{\kappa}_{-\bmf{l}'}}
+\skaco{\tilde{\kappa}_{\bmf{l}}\tilde{\kappa}_{\bmf{l'}}}
\skaco{\tilde{\kappa}_{-\bmf{l}}\tilde{\kappa}_{-\bmf{l}'}}
+\skaco{\tilde{\kappa}_{\bmf{l}}\tilde{\kappa}_{-\bmf{l'}}}
\skaco{\tilde{\kappa}_{-\bmf{l}}\tilde{\kappa}_{\bmf{l}'}}
+\skaco{\tilde{\kappa}_{\bmf{l}}\tilde{\kappa}_{-\bmf{l'}}
\tilde{\kappa}_{-\bmf{l}}\tilde{\kappa}_{\bmf{l}'}}_c
\nonumber\\
&&\hspace{-2em}
=\Omega_{\rm s}^2P_\kappa(l)P_\kappa(l')
+\Omega_{\rm s}^2P_\kappa(l)\delta^K_{\bmf{l}+\bmf{l}'}P_\kappa(l')\delta^K_{\bmf{l}+\bmf{l}'}
+
\Omega_{\rm s}^2P_\kappa(l)\delta^K_{\bmf{l}-\bmf{l}'}
P_\kappa(l')\delta^K_{\bmf{l}-\bmf{l}'}+
\Omega_{\rm s}T_\kappa(\bm{l},-\bm{l},\bm{l}',-\bm{l}').
\end{eqnarray}
Inserting this equation into Eq.~(\ref{eqn:pcovdef}) gives
\begin{eqnarray}
[\bm{C}^g]_{ll'}&=&
\frac{1}{N_p(l)}\frac{1}{N_p(l')}\sum_{\bmf{l}; l\in l_b}\sum_{\bmf{l}'; l'\in l'_b}
\left[
P_\kappa(l)\delta^K_{\bmf{l}+\bmf{l}'}P_\kappa(l')\delta^K_{\bmf{l}+\bmf{l}'}
+P_\kappa(l)\delta^K_{\bmf{l}-\bmf{l}'}P_\kappa(l')\delta^K_{\bmf{l}-\bmf{l}'}
\right]+\frac{1}{N_p(l)N_p(l')\Omega_{\rm s}}\sum_{\bm{l}; l\in l_b}
\sum_{\bm{l}'; l'\in l'_b}T_\kappa
\nonumber\\
&=&\frac{1}{N_p(l)}\frac{1}{N_p(l')}\sum_{\bmf{l}; l\in l_b}\sum_{\bmf{l}'; l'\in l_b}
2P^2_\kappa(l)\delta^K_{\bmf{l}+\bmf{l}'}
+\frac{1}{N_p(l)N_p(l')\Omega_{\rm s}}\sum_{\bm{l}; l\in l_b}
\sum_{\bm{l}'; l'\in l'_b}T_\kappa(\bm{l},-\bm{l},\bm{l}',-\bm{l}')
\nonumber\\
&=&\frac{1}{N_p(l)}\frac{1}{N_p(l')}\sum_{\bmf{l}; l\in l_b}
2P^2_\kappa(l)\delta^K_{ll'}
+\frac{1}{N_p(l)N_p(l')\Omega_{\rm s}}\sum_{\bm{l}; l\in l_b}
\sum_{\bm{l}'; l'\in l'_b}T_\kappa(\bm{l},-\bm{l},\bm{l}',-\bm{l}')\nonumber\\
&\approx&\frac{1}{N_p(l)}\frac{1}{N_p(l')}2P^2_\kappa(l)\delta^K_{ll'}
\sum_{\bmf{l}; |\bm{l}|\in l_b}
+\frac{1}{N_p(l)N_p(l')\Omega_{\rm s}}\sum_{\bm{l}; l\in l_b}
\sum_{\bm{l}'; l'\in l'_b}T_\kappa(\bm{l},-\bm{l},\bm{l}',-\bm{l}')\nonumber\\
&\approx &
\frac{\delta^K_{ll'}}
{l\delta lf_{\rm sky}}
P^2_{\kappa}(l)
+\frac{1}{4\pi f_{\rm sky}}
\int_{|\tilde{\bm{l}}|\in l}\!\!\frac{d^2\tilde{\bmf{l}}}{A(l)}
\int_{|\tilde{\bm{l}}'|\in
l'}\!\!\frac{d^2\tilde{\bmf{l}}'}{A(l')}
T_{\kappa}(\tilde{\bmf{l}},-\tilde{\bmf{l}},\tilde{\bmf{l}}',-\tilde{\bmf{l}}'),
\label{eqn:ps-cov}
\end{eqnarray}
where $A(l)\equiv \int_{|\bm{l}|\in l}d^2\bm{l}\approx 2\pi l\delta l$.
In the third equality for the first term calculation, we have replaced
the Kronecker-type delta function for vectors,
$\delta^K_{\bm{l}+\bm{l}'}$, with the delta function for scalar,
$\delta^K_{ll'}$, because the first term is non-vanishing only if two
multipoles $l$ and $l'$ are same to within the bin widths. In the fifth
line, we have used the integral form for the second term rather than
the summation form for notational simplicity.  The first term of the
covariance represents the Gaussian errors where the power spectrum of
different multipoles are independent, while the second term represents
the non-Gaussian errors to describe correlations between the power
spectra in different multipole bins.  Extending Eq.~(\ref{eqn:ps-cov})
to the tomographic case for source redshift distribution gives
Eq.~(\ref{eqn:pscov}) in main text. 
The equation (\ref{eqn:ps-cov}) is
equivalent to the expression  used in \cite{coorayhu01}
when $l,l'\gg
1$.

\subsection{Cross-covariance between the cluster number counts and
the lensing power spectrum}
\label{appendix:ccwlcov}

We can now derive the cross-covariance between the cluster counts and
the cosmic shear power spectrum. For illustrative purposes, we consider a
single redshift bin for both the lensing power spectrum measurement and
the cluster counts.  From Eqs.~(\ref{eqn:angcl}) and
(\ref{eqn:psest_appendix}), the cross-covariance is defined as
\begin{eqnarray}
 {\rm Cov}[{\cal N}_{\rm cl},P^{\rm est}_\kappa(l)]&\equiv&
\skaco{{\cal N}_{\rm cl}P^{\rm est}_\kappa(l)}-
NP_\kappa(l)
\nonumber\\
&=&
\kaco{\delta\!N_{\rm cl}(\bmf{\theta})P_\kappa^{\rm est}(l)}
\nonumber\\
&=&\frac{1}{\Omega_{\rm s}^2N_p(l)}
\sum_{\bmf{l}; l\in l_b}\int\!\!d^2\bmf{\theta}W(\bmf{\theta})
\kaco{\tilde{\kappa}_{\bmf{l}}\tilde{\kappa}_{-\bmf{l}}
\delta\!N_{\rm cl}(\bmf{\theta})}
\nonumber\\
&=&\frac{1}{\Omega_{\rm s}^2N_p(l)}\int\!d^2\bmf{\theta}W(\bmf{\theta})
\sum_{\bmf{l};l\in l_b}\sum_{\bmf{l}'}\skaco{\tilde{\kappa}_{\bmf{l}}
\tilde{\kappa}_{-\bmf{l}}\delta\tilde{N}_{\bmf{l}'}}e^{i\bmf{l}'\cdot\bmf{\theta}}.
\label{eqn:covcwldef}
\end{eqnarray}
In the second line on the r.h.s., we introduced the angular number
density fluctuation field of the cluster counts, $\delta\!N_{\rm
cl}(\bm{\theta})$, for convenience for the following discussion. Please
do not confuse $\delta\!N_{\rm cl}(\bm{\theta})$ with the ensemble
average of the average angular number density, $N$. The angular number
density fluctuation field can be expressed in terms of the
three-dimensional number density fluctuation field of clusters defined
in Eq.~(\ref{eqn:ndens}):
\begin{equation}
\delta\! N_{\rm cl}(\bmf{\theta})=\int\!\!d\chi\frac{d^2V}{d\chi d\Omega}\delta\!n_{\rm
 cl}(\chi,\chi\bmf{\theta}).
\end{equation}
In the fourth line on the r.h.s. of Eq.~(\ref{eqn:covcwldef}), we used
the Fourier transform of $\delta\!N_{\rm cl}$ using the discrete Fourier
decomposition for a finite sky survey as discussed around
Eq.~(\ref{eqn:four}):
\begin{equation}
\delta\! N_{\rm cl}(\bmf{\theta})=
\frac{1}{\Omega_{\rm s}}
\sum_{\bmf{l}'}\delta\!\tilde{N}_{\bmf{l}'}e^{i\bmf{l}'
\cdot\bmf{\theta}}.
\end{equation}

As performed in Eq.~(\ref{eqn:psdef}), the ensemble average of
products of the cluster number density fluctuation field and the two
convergence fields, appearing in the last line on the r.h.s. of
Eq.~(\ref{eqn:covcwldef}), can be expressed in terms of the angular
bispectrum defined as
\begin{equation}
\skaco{\tilde{\kappa}_{\bmf{l}}
\tilde{\kappa}_{-\bmf{l}}\delta\tilde{N}_{\bmf{l}'}}\equiv
\Omega_{\rm s}B_{\rm c-wl}(\bm{l}',
\bmf{l},-\bmf{l},)\delta^{K}_{\bmf{l}'}.
\label{eqn:angbispdef}
\end{equation}
Substituting this equation into Eq.~(\ref{eqn:covcwldef}) allows further
simplification of Eq.~(\ref{eqn:covcwldef}) as
\begin{eqnarray}
{\rm Cov}[{\cal N},P^{\rm est}(l)]&=&
\frac{1}{\Omega_{\rm s}N_p(l)
}\int\!d^2\bmf{\theta}W(\bmf{\theta})\sum_{\bmf{l}; l\in l_b} B_{\rm
c-wl}(\bm{l}'=\bm{0},\bmf{l},-\bmf{l}) \nonumber\\
&=&\frac{1}{\Omega_{\rm s}N_p(l)}\sum_{\bmf{l}; l\in l_b} B_{\rm
c-wl}(\bm{l}'=\bm{0},\bmf{l},-\bmf{l}) \nonumber\\
&\approx & \frac{1}{\Omega_{\rm s}}B_{\rm c-wl}(l'=0,l,l),
\label{eqn:covcwldef2}
\end{eqnarray}
where we have used $\int\!\!d^2\bm{\theta}W(\bm{\theta})=1$ and assumed,
in the third line on the r.h.s., that the bispectrum $B_{\rm
c-wl}(\bm{l}'=\bm{0},{\bm{l}},-\bm{l})$ changes little within the
multipole bin width.

Employing Limber's approximation, the angular bispectrum
(\ref{eqn:angbispdef}) can be expressed in terms of the 3D bispectrum
defined by Eq.~(\ref{eqn:bisphalos}) as
\begin{equation}
B_{\rm c-w}(l_1,l_2,l_3)= \int\!\!d\chi~ \frac{d^2V}{d\chi d\Omega}
W_{ g}^2(\chi)\frac{1}{\chi^4} \bar{n}_{\rm cl}(\chi)
B_{c\delta\delta}(k_1=l_1/\chi, k_2=l_2/\chi,k_3=l_3/\chi).
\end{equation}
Inserting this equation into Eq.~(\ref{eqn:covcwldef2}) gives the final
expression for the cross-covariance between the cluster counts and the
lensing power spectrum:
\begin{equation}
{\rm Cov}[{\cal N}, P_\kappa^{\rm est}(l)]=
\frac{1}{\Omega_{\rm s}} \int\!\!d\chi~ \frac{d^2V}{d\chi d\Omega}
W_{g}^2(\chi)\frac{1}{\chi^4} \bar{n}_{\rm cl}(\chi)
B_{c\delta\delta}(k_1=0, k_2=l/\chi,k_3=l/\chi).
\label{eqn:cc-wlcov}
\end{equation}
Note that the cross-covariance scales with the sky coverage as $1/f_{\rm
sky}$.  Further, it would be instructive to explicitly show each of the
1-, 2- and 3-halo term contributions to the covariance.  Inserting
Eq.~(\ref{eqn:bisp1h}) into Eq.~(\ref{eqn:cc-wlcov}), the 1-halo term
contribution to the covariance can be expressed as
\begin{equation}
{\rm Cov}[{\cal N}, P^{\rm est}_\kappa(l)]^{1h}=
\frac{1}{\Omega_{\rm s}}
\int\!\!d\chi~ \frac{d^2V}{d\chi d\Omega}
W_{g}^2(\chi)\frac{1}{\chi^4}
\int\!\!dm~ n(m)\left(\frac{m}{\bar{\rho}_m}\right)^2
S(m) \tilde{u}_m(k=l/\chi)\tilde{u}_m(k=l/\chi).
\label{eqn:cc-wlcov-1h}
\end{equation}
Similarly, from Eq.~(\ref{eqn:bisp2h}), the 2-halo term contribution is
found to be
\begin{equation}
{\rm Cov}[{\cal N}, P^{\rm est}_\kappa(l)]^{2h}=
\frac{2}{\Omega_{\rm s}}
\int\!\!d\chi~ \frac{d^2V}{d\chi d\Omega}
W_{g}^2(\chi)\frac{1}{\chi^4}
\left[
\int\!\!dm~ n(m)S(m)b(m)\frac{m}{\bar{\rho}_m}
\tilde{u}_{m}(k=l/\chi)
\right]^2P_\delta^L(k=l/\chi; \chi),
\label{eqn:cc-wlcov-2h}
\end{equation}
where we have used $P^L(k)\rightarrow 0$ for $k\rightarrow 0$ for the
linear power spectrum as predicted from the inflation motivated
primordial power spectrum.  It is also interesting to find the 3-halo
term contribution to the covariance is vanishing as
\begin{equation}
{\rm Cov}[{\cal N}, P^{\rm est}_\kappa(l)]^{3h}=0,
\label{eqn:cc-wlcov-3h}
\end{equation}
because the perturbation theory bispectrum $B^{\rm
PT}(\bmf{k}_1,\bmf{k}_2,\bmf{k}_3)\rightarrow 0$ for $k_1\rightarrow 0$.

\section{A toy model}
\label{appendix:toy-wl}

In this section we show that the qualitative behavior of the
percentage difference in $S/N$ (plotted in the lower panel of
Figs.~\ref{fig:sn} and \ref{fig:sn-wlsn})
can be recovered using a surprisingly simple toy model.
Although extremely simple it will help us gain intuition for the cause
of the increase and decrease in the percentage difference in $S/N$.

Imagine a simple universe in which all halos exist only at a small range of redshifts
around redshift $z$,
are unclustered ($P^L(k)=0$), and have a profile which is independent of mass
($\tilde{u}^2_m(k=l/\chi)\sim\tilde{u}^2(k=l/\chi)$).
Eq.~(\ref{eqn:cc-Nb}) for cluster counts
may be rewritten as
\begin{equation}
D^c \equiv N = \chi^2 \delta\chi
\int\!\!dm~ S(m) n(m),
\end{equation}
where we assume that the volume element over a small redshift interval
is given by $\int\!\!d\chi (d^2V/d\chi d\Omega)=\chi^2 \delta \chi$, and
consider
a single redshift bin.
Further, from Eq.~(\ref{eqn:p_kappa}) and using the halo model
expression for the 1-halo term of the 3D mass power spectrum (e.g. see
Eq.~[9] in \cite{TJ03a}),
the lensing power spectrum for our simple universe
can be shown to be given by
\begin{equation}
D^g \equiv P_{\kappa}=
W_g^2(\chi)\chi^{-2}\delta\chi \tilde{u}^2
\int\!\!dm~ n(m)\left(\frac{m}{\bar{\rho}_m}\right)^2.
\label{eqn:toy-wl}
\end{equation}
For simplicity we consider observations at a single
$\ell$ using one redshift bin, therefore the lensing power spectrum measurement
is a single number.

To reproduce the results in Fig.~\ref{fig:sn} we also need to calculate the
covariances in this simple universe.
We will assume that there is no galaxy intrinsic ellipticity
$\sigma_{\epsilon}=0$  so that the shot noise term due to intrinsic galaxy shapes is
negligible.
Further we assume that the lensing power spectrum covariance arises only
from
the 1-halo term of the lensing trispectrum in
Eq.~(\ref{eqn:pscov}), or in other words ignore the Gaussian error
contribution (the first term in Eq.~[\ref{eqn:pscov}]).
Substituting the 1-halo term of the lensing trispectrum into
Eq.~(\ref{eqn:pscov}) for this simple universe gives
\begin{equation}
\bm{C}^{g}
=\frac{1}{\Omega_s} W_g^4(\chi)\chi^{-6}\delta\chi
\tilde{u}^4
\int\!\!dm~ n(m)\left(\frac{m}{\bar{\rho}_m}\right)^4.
\label{eqn:toy-wl-1h}
\end{equation}
The cluster count variance is given by the shot noise (see
Eq.~[\ref{eqn:clcov}]):
\begin{equation}
\bm{C}^{c}
=
\frac{1}{\Omega_{\rm s}}\chi^2 \delta\chi \int\!\!dm~ S(m) n(m).
\label{eqn:toy-c-1h}
\end{equation}
The cross-covariance between cluster counts and the lensing power spectrum
is expected from Eq.~(\ref{eqn:cc-wlcov}) to be
\begin{equation}
\bm{C}^{gc}
=\frac{1}{\Omega_{\rm s}}W_g^2(\chi) \chi^{-2}\delta\chi \tilde{u}^2
\int\!\!dm~ S(m)n(m)\left(\frac{m}{\bar{\rho}_m}\right)^2.
\label{eqn:toy-cwl-1h}
\end{equation}
Thus the only ingredient is the mass function weighted
by various powers of the mass, and by the cluster selection function,
the lensing efficiency and the volume element.

\begin{figure}[th]
\centerline{\psfig{file=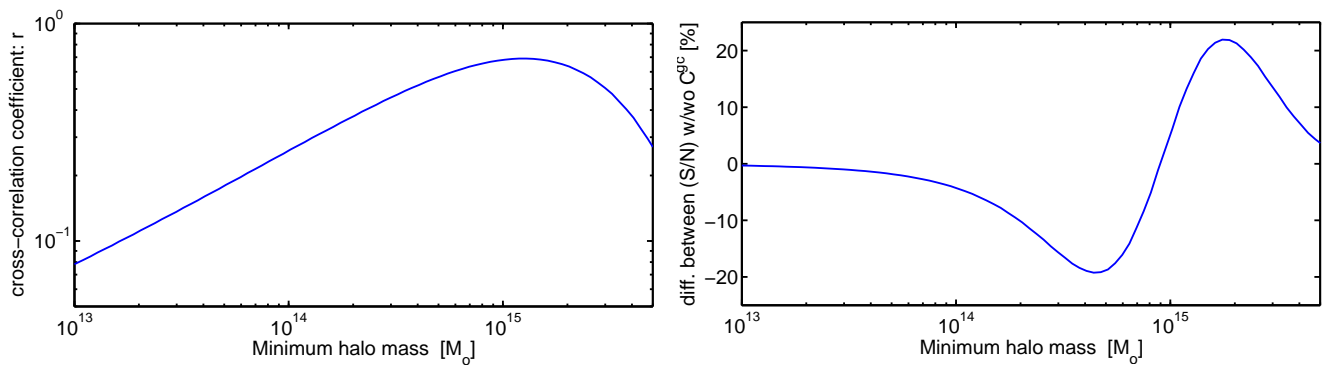, width=17.5cm}}
\caption{{\em Left panel}:
Correlation coefficient $r$ between cluster counts and the lensing power spectrum
for the simple toy model containing only the halo
mass function and mass weighting. {\em Right panel}:
Percentage difference in $S/N$ (to be compared with the lower panel of
Fig.~\ref{fig:sn}) for the simple toy model containing only the halo
mass function and mass weighting.
}
\label{fig:toy_R}
\end{figure}

The correlation coefficient
between the cluster counts and lensing power spectrum
is given as before as
\begin{equation}
r = \frac{ \bm{C}^{gc} }{ \sqrt{ \bm{C}^{c} \bm{C}^{g}}}.
\end{equation}
Note that all the prefactors in Eqs.~(\ref{eqn:toy-wl-1h}), (\ref{eqn:toy-c-1h}) and
(\ref{eqn:toy-cwl-1h}) appearing in front of the
halo mass integral drop out
and therefore the results shown in this Appendix are independent of redshift and multipole.
This is plotted in the left panel of Fig.~\ref{fig:toy_R}
and may be compared to the lower panel of Fig.~\ref{fig:r-cov} for
the full treatment.
The shape is remarkably similar to that for the full treatment,
implying that this simple model containing the different weightings
of the mass function captures the essence of the
complementarity.
The correlation peaks at around $10^{15} M_{\odot}$ and decreases at low minimum
cluster masses.
It makes sense that the cluster counts and lensing are correlated at high
minimum cluster masses, because the mass weighting in the toy lensing
power spectrum is similar to the mass cut in the cluster counts:
they are both dominated by high mass clusters.
They become less correlated at lower minimum masses because the cluster
counts are dominated by low mass clusters that contribute less to the
lensing power spectrum.

The size of the correlation is larger for this simple model
than for the full treatment. This makes sense because the full
model contains several ingredients that will reduce the correlation
including: the contributions of the Gaussian errors and
shot noise to the lensing power spectrum covariance (e.g., compare thick and thin lines in
the lower panel of Fig.~\ref{fig:r-cov});
halos at a range of redshifts which will be weighted differently
by the lensing power spectra and cluster counts;
and terms involving more than one halo at a time.
However an \emph{even} simpler toy model, in which all
halos in the universe have the same mass,
would make lensing power
spectra and cluster counts 100 per cent correlated.
We see that simply including the mass weighting $\propto m^2$
for the lensing power spectra stops a complete redundancy of
information and starts to explain how these seemingly similar
probes can be so complementary.

In this simple model we have only one data point from cluster
counts $D^c$ and one data point from lensing $D^g$ (since we have only
one redshift bin for each, and we are considering a single
wavenumber $l$).
Therefore
we can write the signal-to-noise in terms of the correlation coefficient
\begin{equation}
\left(\frac{S}{N}\right)^2 = \frac{1}{(1-r^2)} \left(
\frac{D^{g \,2}}{\bm{C}^g} +
\frac{D^{c \,2}}{\bm{C}^c} -
\frac{2 D^{g}D^{c}  }{\bm{C}^{gc}} r \right).
\end{equation}
All the terms scale with ($\Omega_{\rm s}\chi^2\delta\chi$),
the comoving volume of the redshift interval
considered.
We use a redshift slice at
$z=0.3$ of thickness $0.1$ for illustration.

To reproduce the plot of percentage difference in $(S/N)$
we need to compare this to the $(S/N)$ when the covariance
is not taken into account, found by setting $r=0$ in the above.
The factor $1/(1-r^2)$ is close to unity when the correlation
coefficient $r$ is small,
but when the correlation is strong ($r\sim1$)
it gets much larger. This causes the $(S/N)$ to be larger
when the covariance is included than when it is not included and gives
rise to the peak in the right hand panel of Fig.~\ref{fig:toy_R}.
The final term in the round brackets
causes a decrease in $S/N$, relative to the case where
no covariance is included.
This is important especially
when the correlation is small and the factor  $1/(1-r^2)$ is
unimportant.
This explains the dip in right hand panel of Fig.~\ref{fig:toy_R}.
In summary: the fact that Fig.~\ref{fig:toy_R} is qualitatively similar
to the lower panel of Fig.~\ref{fig:sn} suggests that the peak and dip can
be explained just in terms of the different mass weighting of cluster counts
and the lensing power spectrum.

\end{document}